\documentclass[%
aip,amsmath,amssymb,reprint]{revtex4-1}

\usepackage{graphicx}
\usepackage{dcolumn}
\usepackage{bm}

\usepackage[utf8]{inputenc}
\usepackage[T1]{fontenc}
\usepackage{mathptmx}
\usepackage{gensymb}
\usepackage{textcomp}
\usepackage[usenames,dvipsnames]{color}

\begin{document}

\preprint{AIP/123-QED}

\title{Deterministic and stochastic aspects of current-induced magnetization reversal in perpendicular nanomagnets}


\author{G. Sala}
\email{giacomo.sala@mat.ethz.ch}
\affiliation{Department of Materials, ETH Zurich, 8093 Zurich, Switzerland}
\author{J. Meyer}
\affiliation{Department of Materials, ETH Zurich, 8093 Zurich, Switzerland}
\author{A. Flechsig}
\affiliation{Department of Materials, ETH Zurich, 8093 Zurich, Switzerland}
\author{L. Gabriel}
\affiliation{Department of Materials, ETH Zurich, 8093 Zurich, Switzerland}
\author{P. Gambardella}
\email{pietro.gambardella@mat.ethz.ch}
\affiliation{Department of Materials, ETH Zurich, 8093 Zurich, Switzerland}

\begin{abstract}
We study the incubation and transition times that characterize the magnetization switching induced by spin-orbit torques in nanomagnets with perpendicular anisotropy. We present a phenomenological model to interpret the dependence of the incubation time on the amplitude of the voltage pulse and assisting magnetic field, and estimate the volume of the seed domain that triggers the switching. Our measurements evidence a correlation between the incubation and transition times that is mediated by the temperature variation during the electric pulse. In addition, we discuss the stochastic distributions of the two times in terms of the energy barriers opposing the nucleation and expansion of the seed domain. We propose two models based on the log-normal and gamma functions to account for the different origin of the variability of the incubation and transition times, which are associated with a single nucleation barrier and multiple pinning sites, respectively.
\end{abstract}

\maketitle

\textit{The original manuscript has been published with DOI 10.1103/PhysRevB.107.214447 by the American Physical Society, which ows the copyright of the published article. Publication on a free-access e-print server of files prepared and formatted by the authors is allowed by the American Physical Society. The authors own the copyright of these files.}

\vspace{0.5cm}

\section{Introduction}
Current-induced magnetization switching by either spin-transfer torque (STT) [\onlinecite{Myers1999,Wegrowe1999,Stiles2002}] or spin-orbit torque (SOT) [\onlinecite{Miron2011,Liu2012a,Manchon2019}] provides the most effective means to achieve electrical control of nanoscale magnetic devices [\onlinecite{Khvalkovskiy2013a,Krizakova2022}]. Although the switching can be described theoretically by macrospin models [\onlinecite{Sun2000,Lee2013}], the magnetization reversal caused by STT and SOT is in general a complex and non-uniform process that involves the nucleation and expansion of magnetic domains [\onlinecite{Bernstein2011,Baumgartner2017}]. This mechanism prevails over the pure macrospin dynamics even in devices as small as few tens of nm because it is more energetically favorable than the coherent rotation of all the magnetic moments [\onlinecite{Sun2011,Chaves-OFlynn2015,BeikMohammadi2021,Bouquin2021}]. Time-resolved measurements in magnetic tunnel junctions [\onlinecite{Devolder2008,Tomita2008,Grimaldi2020,Krizakova2020,Bouquin2021}] and Hall crosses [\onlinecite{Sala2021}] have revealed that the non-uniform switching comprises two phases at the ns and sub-ns timescale: an initial waiting time $t_0$ during which the magnetization is at rest and the actual transition of duration $\Delta t$. The physical mechanisms underlying $t_0$ and $\Delta t$ are similar but not identical in the STT and SOT scenarios.

In the STT-driven dynamics, the magnetization is switched by the torque exerted by the spin-polarized current originating from a reference ferromagnetic layer. The initial waiting time $t_0$ results from the combination of two factors. First, if the electronic spin polarization and the magnetization are collinear, the torque is initially zero and stochastic thermal fluctuations are necessary to trigger the reversal [\onlinecite{Bultynck2018}]. Second, the formation of the seed domain that initiates the reversal always requires to overcome the energy barrier created by the effective magnetic anisotropy. The duration of the nucleation phase is, therefore, influenced by the temperature increase caused by Joule heating, the magnetic field $B$, and the amplitude of the electric pulses (current density $j$ or voltage $V_{\textrm{P}}$) that drive the dynamics. In this case, $t_0$ is modeled by an Arrhenius-type law with an energy barrier $\Delta U$
\begin{equation}
t_0 = \tau e^{\frac{\Delta U} {k_{\textrm{B}}T} }, 
\label{eq:basic_t0}
\end{equation}
which describes thermally activated processes [\onlinecite{Wegrowe1999,Koch2004,Li2004,Bedau2010,Bultynck2018,Desplat2020}]. Here, $\tau^{-1}$ is the attempt frequency and $k_{\textrm{B}} T$ is the thermal energy. $\Delta U = \Delta U(V_{\textrm{P}}, B)$ generally depends on both $V_{\textrm{P}}$ and $B$ such that higher pulse amplitudes and/or larger magnetic fields result in a strong decrease of the nucleation time [\onlinecite{Iga2012,Hahn2016,Bultynck2018,Grimaldi2020}]. The ensuing domain expansion determines the transition time $\Delta t$. Yet, the details of this process depend on the magnetic properties and dimension of the specific sample under investigation as well as the pulse amplitude [\onlinecite{Hahn2016,Visscher2016,Bultynck2018}]. The nucleation can take place at the device edge, and the magnetization reversal can be driven by the motion of a single domain wall across the device. In this case, $\Delta t$ is inversely proportional to the domain wall speed [\onlinecite{Hahn2016,Devolder2018,Bouquin2021}]. However, theory shows that the switching may also be triggered by magnetostatic instabilities in which different device regions are perturbed differently by the STT [\onlinecite{Munira2015,Visscher2016,BeikMohammadi2021}]. Under these circumstances, $t_0$ and $\Delta t$ may not be associated with two distinct processes.

\begin{figure}
\includegraphics[scale=1]{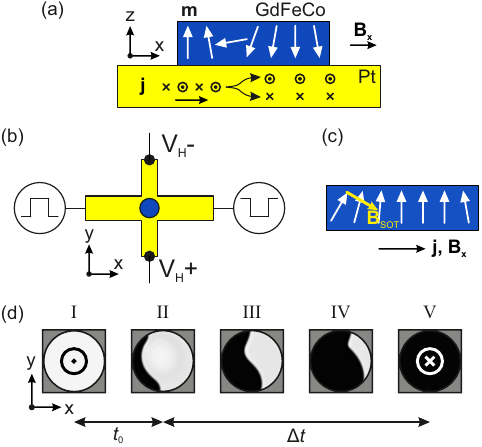}
\caption{(a) Schematic side view of the device. A perpendicularly-magnetized GdFeCo dot (blue) is fabricated on a Pt Hall cross (yellow). Current-induced SOT cause the magnetization switching by domain wall motion in the presence of a magnetic field $\mathbf{B}_x$ collinear to the current direction $\mathbf{j}$. (b) Schematic top view of the device. The detection of the anomalous Hall voltage $V_{\textrm{H}}$ during the injection of ns-long electric pulses allows us to track the magnetization switching in real time [\onlinecite{Sala2021}]. (c) Schematic side view of the magnetization profile before the nucleation of a domain when the current $\mathbf{j}$ and magnetic field $\mathbf{B}_x$ are oriented along $+x$. In Pt/ferromagnet and Pt/ferrimagnet bilayers, the DMI favors inward canting of the magnetic moment when the magnetization is on average pointing up. This canting and the global tilt towards $+x$ caused by $\mathbf{B}$ assist the nucleation of the seed domain by the SOT field $\mathbf{B}_{\textrm{SOT}}$ on the left side. When the magnetization is pointing down, the same combination of current and field leads to the nucleation on the right side (not shown). (d) Schematic of the SOT-induced switching of perpendicular magnets as a function of time (defined by the labels I-V). The black-white contrast indicates the sign of the out-of-plane component of the magnetization ($m_z$). The switching is driven by the nucleation of a domain at the device edge and the ensuing domain wall motion. $t_0$ and $\Delta t$ are the duration of the two phases, respectively. The sketches in (c) and (d) are adapted from Ref.~[\onlinecite{Baumgartner2017}]}.
\label{fig:fig1}
\end{figure}

In the SOT scenario, a spin current is injected into the ferromagnet from an adjacent non-magnetic layer as a consequence of the spin Hall effect and spin-orbit scattering occurring in the non-magnet [\onlinecite{Manchon2019,Krizakova2022}] (Fig. \ref{fig:fig1}a). Therefore, the spin polarization and the magnetization are orthogonal in perpendicular magnets, and no initial latency is expected [\onlinecite{Bedau2010,Liu2014b,Garello2014}]. However, recent experiments have shown that a finite $t_0$ exists also in the SOT-induced dynamics near the critical switching threshold, because the strong energy barrier due to the perpendicular magnetic anisotropy can be more easily overcome with thermal assistance [\onlinecite{Grimaldi2020,Krizakova2020,Sala2021,Cai2021}]. The similarity between the STT and SOT dynamics suggests that the nucleation time may be described by an expression similar to Eq. \ref{eq:basic_t0} even in the SOT scenario. However, this possibility and the validity of statistical models that capture the stochastic aspects of SOT switching have not been tested thus far. Additionally, the dependence of $t_0$ on $V_{\textrm{P}}$ and $B$ (applied parallel to the current direction) in the SOT case has never been investigated quantitatively. In the SOT-driven switching of perpendicular magnets, the nucleation of the seed domain usually happens at the device edge because of the combination of SOT, Dzyaloshinkii-Moriya interaction (DMI), and external magnetic field [\onlinecite{Mikuszeit2015,Martinez2015,Baumgartner2017,Sala2022,Pizzini2014}] (see Fig. \ref{fig:fig1}c,d). Thus, in the SOT scenario, $\Delta t$ is usually identified with the duration of the domain wall displacement between opposite device edges [\onlinecite{Mikuszeit2015,Martinez2015,Baumgartner2017,Sala2022}]. 

This interpretation remains valid as long as domains can be clearly identified. However, the magnetization reversal in small devices assumes a mixed character that is neither pure macrospin nor domain-driven [\onlinecite{Krizakova2022}]. In this crossover regime, $t_0$ and $\Delta t$ are ill-defined and cannot be associated with distinct physical processes. This happens in devices with dimensions comparable to the domain wall width, typically in the order of 10-50 nm in thin films with perpendicular magnetization [\onlinecite{Chaves-OFlynn2015}]. Further, the switching dynamics is expected to change if ps-long current pulses are used to drive the magnetization out-of-equilibrium by reducing the saturation magnetization and magnetic anisotropy at a faster rate than the precessional motion [\onlinecite{Jhuria2020}]. However, the exact magnetic dynamics at these timescales and the relative importance of torques and Joule heating remain unclear [\onlinecite{Yang2017,Wilson2017}].

These considerations show that temperature plays a two-fold role in both the STT- and SOT-driven dynamics. On the one hand, the temperature increase causes a deterministic variation of the magnetic parameters and, hence, of $t_0$ and $\Delta t$. On the other hand, thermal fluctuations lead to a distribution of the nucleation and transition times. Indeed, non-reproducible dynamics have been observed by time-resolved single-shot measurements of the STT-induced switching [\onlinecite{Tomita2008,Bultynck2018,Bouquin2021,Cui2010,Zhao2012,Grimaldi2020,Krizakova2020,Krizakova2021,Cai2021}] and, recently, of the SOT-induced reversal [\onlinecite{Grimaldi2020,Krizakova2020,Sala2021,Krizakova2021,Cai2021}]. These measurements show that both $t_0$ and $\Delta t$ present in general skewed statistical distributions [\onlinecite{Cui2010,Iga2012,Zhao2012,Bouquin2021,Grimaldi2020,Krizakova2020,Sala2021}]. The stochasticity of the STT-driven dynamics has also been analyzed theoretically in a few publications [\onlinecite{Diao2007,Vincent2015,Moon2018,Siracusano2018,DAquino2020,Shukla2020}], but there has been little synergy between the experimental and theoretical approaches. At the experimental level, the choice of a specific distribution to fit the data has been seldom guided by physical considerations. At the theoretical level, exact but rather complex and mathematically non-explicit distribution functions have been proposed [\onlinecite{Diao2007,Vincent2015,Moon2018,Siracusano2018,DAquino2020,Shukla2020}], which prevent an intuitive understanding of the underlying physical processes and limit the applicability of the derived equations. In addition, most theoretical models rest on the assumption of coherent macrospin dynamics, which is rarely fulfilled.


Here, we provide a simple phenomenological interpretation of the deterministic and stochastic dynamics induced by SOT in perpendicular magnets. We have performed systematic time-resolved Hall measurements of the SOT-induced magnetization switching in ferrimagnetic Pt/GdFeCo dots with perpendicular magnetic anisotropy [\onlinecite{Sala2021}] and have analyzed the influence of the pulse amplitude $V_{\textrm{P}}$, assisting in-plane field $B_x$, and device size on the mean $t_0$ and $\Delta t$. We propose a phenomenological model to describe the dependence of $t_0$ on the SOT amplitude, magnetic field, and self-heating. This model validates the qualitative interpretation of the SOT switching as a thermally-assisted process, as proposed in previous works [\onlinecite{Grimaldi2020,Krizakova2020,Krizakova2021,Cai2021,Sala2021}], and allows us to estimate the volume of the seed domain that initiates the reversal. Our data further evidence a clear correlation between $t_0$ and $\Delta t$ that was not identified before and that we attribute to the temperature variation during the electric pulse. Finally, we consider the statistical distributions of $t_0$ and $\Delta t$ and propose two physically intuitive models based on the log-normal and gamma functions to interpret the stochasticity of the two times. These models suggest a different origin of the stochastical variations of $t_0$ and $\Delta t$, which we associate with a single energy barrier and multiple pinning sites, respectively. Although our analysis focuses on SOT-induced switching, our considerations are general and apply also to the magnetization reversal induced by STT. They are also equally valid for ferro- and ferrimagnets because our models are independent of the ferrimagnetic order of GdFeCo. The only assumption that we make is that switching occurs by the nucleation and expansion of domains.

\section{Experiment}

\begin{figure}
\includegraphics[scale=1]{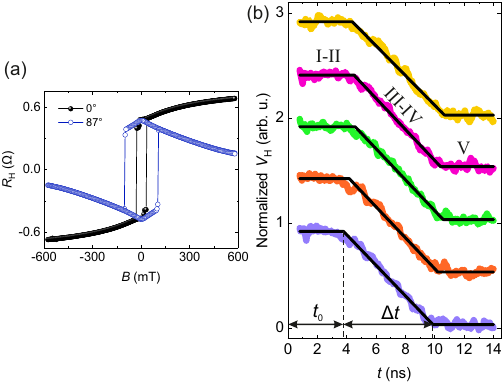}
\caption{(a) Out-of-plane magnetization of Gd$_{29}$Fe$_{64}$Co$_{7}$ measured by the anomalous Hall effect during a magnetic field sweep in the $xz$ plane at a polar angle $\theta_B = 0^{\circ}$ and $\theta_B = 87^{\circ}$ from the normal direction. (b) Exemplary single-shot measurements of the up-down magnetization reversal in a 5-\textmu m-wide device switched by electric pulses with $V_{\textrm{P}}=30$~V and 15 ns duration and $B=50$~mT applied parallel to the current direction. The straight lines are piece-wise linear fits of the quiescent, reversal, and equilibrium phase. The voltage traces are offset for clarity, and the times $t_0$ and $\Delta t$ of the first switching event are indicated. The labels I-V refer to Fig. \ref{fig:fig1}d.}
\label{fig:fig3}
\end{figure}

Figure \ref{fig:fig1}a,b shows the structure of our devices and experimental scheme. A 15-nm-thick Gd$_{29}$Fe$_{64}$Co$_{7}$ pillar is fabricated above a 5-nm-thick Pt Hall cross by optical lithography and ion-beam etching. These devices may be considered equivalent to the lower half of a top-pinned magnetic tunnel junction. Gd$_{29}$Fe$_{64}$Co$_{7}$ is a ferrimagnet with room-temperature net magnetization $M = M_{\textrm{FeCo}}-M_{\textrm{Gd}}$ dominated by the FeCo sublattice and with perpendicular magnetic anisotropy, as indicated by the hysteresis loops in Fig. \ref{fig:fig3}a. We chose a rare-earth, transition-metal ferrimagnet because of the large anomalous Hall effect of these materials [\onlinecite{Mimura1978}], which facilitates time-resolved measurements in Hall crosses [\onlinecite{Sala2021,Krishnaswamy2022,Sala2022b}]. However, the analysis and considerations presented below are independent of the antiferromagnetic order of ferrimagnetic materials and, hence, apply also to ferromagnets. In addition, in the following, we consider explicitly a pillar with 5 \textmu m diameter, but we have verified that our considerations are valid in devices with smaller dimensions (1, 3 and 4 \textmu m) as well as different composition. In these samples, we have performed time-resolved single-shot Hall measurements of the magnetization switching induced by SOT, as described in detail in Ref. [\onlinecite{Sala2021}]. In short, two electric pulses with the same amplitude but opposite polarity are injected simultaneously into the Pt Hall cross (Fig. \ref{fig:fig1}b). The electric current exerts SOT on the magnetization and, at the same time, generates the anomalous Hall voltage $V_{\textrm{H}}$ proportional to the out-of-plane magnetization. Therefore, the real-time detection of $V_{\textrm{H}}$ allows us to track with 50 ps temporal resolution the orientation of the magnetization excited by SOT. Exemplary time-resolved measurements performed with this approach are presented in Fig. \ref{fig:fig3}b. Here, the temporal variation of the normalized voltage $V_{\textrm{H}}$ indicates that the reversal comprises three phases: the initial quiescent phase of duration $t_0$, the transition during the time $\Delta t$, and the final equilibrium phase. As discussed before [\onlinecite{Grimaldi2020,Sala2021}], we assume that $t_0$ and $\Delta t$ are the times required to nucleate a domain and move the domain wall across the device, respectively (Fig. \ref{fig:fig1}d). Both $t_0$ and $\Delta t$ are expected to depend on the pulse amplitude $V_{\textrm{P}}$ and magnetic field $B_x$, which is applied collinear to the current direction to ensure the deterministic magnetization reversal [\onlinecite{Manchon2019}]. For each combination of these two parameters, we have taken a set of 250 measurements that, after fitting a piece-wise linear function to the normalized voltage traces (see Fig. \ref{fig:fig3}b), yielded the distributions of the nucleation and transition times. Hereafter, $t_0$ and $\Delta t$ will stand for the statistical means of the respective distributions.

\section[Dependence of $t_0$ and $\Delta t$ on the pulse amplitude and magnetic field]{Dependence of $t_0$ and $\Delta t$ on the pulse amplitude and magnetic field}

Figure \ref{fig:fig4}a-d shows the dependence of $t_0$ and $\Delta  t$ on the pulse amplitude and in-plane magnetic field. Both times decrease upon increasing either parameter, and so do their standard deviations, as also observed in previous works [\onlinecite{Grimaldi2020,Krizakova2020,Sala2021}]. Therefore, the dynamics becomes faster and more reproducible when overdriven. Moreover, for a given pair of current density and magnetic field, both $t_0$ and $\Delta t$ decrease with the device diameter $d$, as shown in Fig. \ref{fig:fig4}e,f. 
The linear scaling of $\Delta  t$ with $d$ corroborates the assumption that the transition time is inversely proportional to the domain wall speed. Indeed, the dependence of $\Delta  t$ on $B_x$ and $V_{\textrm{P}}$ agrees qualitatively with predictions of the dynamics of domain walls driven by SOT in the presence of a magnetic field [\onlinecite{Haazen2013,Emori2013,Emori2014,Martinez2014}]. In these models, a more intense current exerts stronger torques on the magnetization of the domain wall and speeds up its motion. However, since SOT also push the magnetization $\mathbf{m}$ away from the current direction, the domain wall velocity tends to saturate [\onlinecite{Martinez2014,Yang2015}]. The saturation occurs because the damping-like SOT is proportional to the product $m_x j_{\textrm{c}} \sim m_xV_{\textrm{P}}$, which becomes constant in the high current limit. This explains why $\Delta  t$ approaches an asymptotic value at large pulse amplitudes. As shown in Fig. \ref{fig:fig4}d, the asymptote is influenced by the magnetic field because the latter reorients the saturation magnetization along $x$, thereby increasing $m_x$. The same argument explains why $\Delta t$ decreases as the field is increased at constant pulse amplitude (Fig. \ref{fig:fig4}c).

The decrease of $\Delta  t$ with $d$ is beneficial to the scalability and speed of spintronic devices. For example, we extrapolate from Fig. \ref{fig:fig4}e that the transition time of a 100-nm-wide device could be of the order of 120 ps for a current density of $j_{\textrm{c}} = 9\cdot10^{11}$ A/m$^2$ and a magnetic field $B = 60$ mT. Thus, the switching could be accomplished with pulses as short as $\approx 150$ ps. Decreasing further the diameter would reduce the minimum pulse length necessary to switch the magnetization, but a simple estimate of $\Delta  t$ for $d \ll 100$ nm  cannot be drawn from our data because of the onset of coherent dynamics in small devices.

\subsection{Phenomenological model of the dependence of $t_0$ on the pulse amplitude and magnetic field}

\begin{figure}
\includegraphics[scale=1]{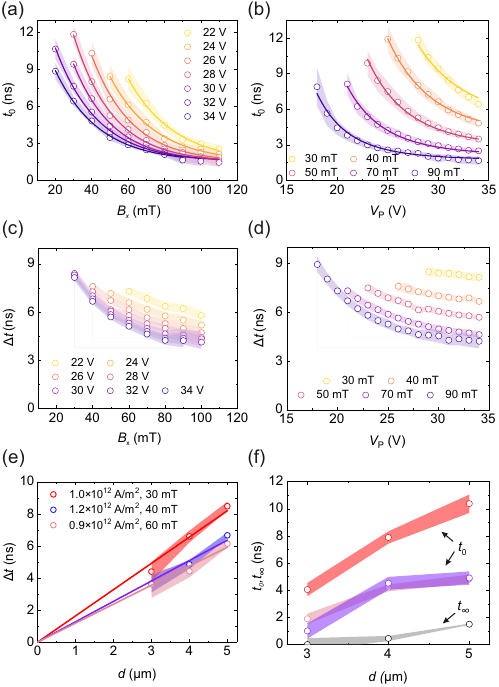}
\caption{(a), (b) Dependence of the mean nucleation time $t_0$ on $B_x$ and $V_{\textrm{P}}$, respectively. The shaded areas define the standard deviation of the distributions of $t_0$. The lines are fits with shared parameters of Eq. \ref{eq:expFit_1} to the data. (c), (d) Same as (a), (b) for the transition time $\Delta  t$. (e) Scaling of $\Delta  t$ with the device diameter $d$ for three different combinations of current density and magnetic field. The lines are fit to the data with zero intercept. (f) Scaling of $t_0$ and $t_{\infty}$ with $d$. The legend of $t_0$ is the same as that of $\Delta t$ in (e). The gray shaded area of $t_{\infty}$ represents the uncertainty in the fit of Eq. \ref{eq:expFit_1} to the field dependence of $t_0$.}
\label{fig:fig4}
\end{figure}

As mentioned before, the dependence of $t_0$ on $V_{\textrm{P}}$ and $B_x$ has been discussed in the framework of the STT-induced magnetization reversal [\onlinecite{Wegrowe1999,Koch2004,Li2004,Bultynck2018,Desplat2020}], but these models have not been extended to SOT yet. To fill the gap, we propose here a simple phenomenological model that builds upon the interpretation of the SOT-induced switching as a thermally-assisted process [\onlinecite{Lee2014,Grimaldi2020,Krizakova2022}]. We argue that the non-zero nucleation time is determined by the energy barrier $\Delta U$ that hinders the formation of a reversed domain and its domain wall. This barrier is lowered by the synergetic action of SOT, magnetic field, DMI, and Joule heating [\onlinecite{Mikuszeit2015,Baumgartner2017,Grimaldi2020,Sala2022}]. The relationship between $t_0$ and $\Delta U$ is captured by an Arrhenius-type law similar to Eq. \ref{eq:basic_t0}, as typical of thermal processes and similar to the STT-induced dynamics [\onlinecite{Li2004,LoConte2014}]. The spread of $t_0$ is then the result of thermal fluctuations, as discussed later. In this scenario, $\Delta U$ equals the energy cost of rotating in plane some of the magnetic moments at the edge of the dot, where the reversal starts [\onlinecite{Mikuszeit2015,Baumgartner2017}] (see Figs. \ref{fig:fig1}d and \ref{fig:fig5}a), and may be expressed as follows:
\begin{align}
\Delta U = \Delta U_B + \Delta U_{\textrm{ex}} + \Delta U_{\textrm{k}} + \Delta U_{\textrm{DMI}} = \Delta U_B + \Delta U_0 .
\end{align}
Here, we consider the contributions of the Zeeman energy $\Delta U_B$, exchange energy $\Delta U_{\textrm{ex}}$, effective anisotropy energy $\Delta U_{\textrm{k}}$, and DMI $\Delta U_{\textrm{DMI}}$, and assume that $\Delta U_0 = \Delta U_{\textrm{ex}} + \Delta U_{\textrm{k}} + \Delta U_{\textrm{DMI}}$ is independent of the applied magnetic field. This assumption does not affect the validity of our model, which could be refined to include, for example, the field-dependence of the effective anisotropy. The independence of the exchange, anisotropy, and DMI energy of the magnetic field is consistent with the so-called "droplet model" [\onlinecite{Pizzini2014,Kim2017d}]. The present analysis can be considered as a reformulation of this model with no explicit reference to the analytical form of $\Delta U_0$. In general, we expect a decrease of the magnetization and anisotropy caused by Joule heating [\onlinecite{Grimaldi2020}] and, hence, a dependence of $\Delta U$ on the applied voltage. Further, we postulate that the SOT itself contributes to the reduction of the energy barrier. In principle, a well-defined energy associated with the torques does not exist because the corresponding effective fields cannot be derived from an energy functional [\onlinecite{Li2004,Manchon2019}]. However, it is still possible to conceive an effective torque contribution to the energy barrier, as it is often assumed in the STT theory [\onlinecite{Koch2004,Li2004,Bedau2010,LoConte2014,Munira2015,Bultynck2018}]. The direct (via the SOT) and indirect (via Joule heating) effects of the pulse amplitude on the energy barrier are unknown and may not have the same functional dependence on $V_{\textrm{P}}$. Therefore, we assume for simplicity that, to the lowest order, $\Delta U$ can be corrected by a term proportional to $V_{\textrm{P}}$, i.e., $\Delta U \rightarrow \Delta U - \gamma V_{\textrm{P}}$.
Then, $t_0$ may be calculated as
\begin{align}
t_0 (V_{\textrm{P}}, B_x) = \tau(V_{\textrm{P}}) e^{\frac{\Delta U (B_x) - \gamma V_{\textrm{P}}}{k_{\textrm{B}} T+\eta V_{\textrm{P}}^2}},
\end{align}
where $\eta V_{\textrm{P}}^2$ accounts for the temperature increase caused by the current. $\tau(V_{\textrm{P}})$ is a characteristic time that could depend on both $V_{\textrm{P}}$ and $B_x$ [\onlinecite{Wegrowe1999,Desplat2020}]. Since the dependence of $t_0$ on $B_x$ resembles a simple exponential (see Fig. \ref{fig:fig4}a), we assume that the effect of the field on $\tau$, if any, is negligible. This assumption is corroborated \textit{a posteriori} by the good quality of the fits in Fig. \ref{fig:fig4}a, as discussed later. On the other hand, the data in Fig. \ref{fig:fig4}b do not exclude that $\tau$ might depend on the voltage, so that we keep the notation $\tau(V_{\textrm{P}})$.

A final point to consider is the finite asymptotic value of $t_0$ when $B_x, V_{\textrm{P}} \rightarrow \infty$. We do not have a conclusive explanation for the existence of a finite intrinsic nucleation time $t_{\infty}$, which is of the order of 1.5 ns in Fig. \ref{fig:fig4}a-b. However, we have observed that $t_{\infty}$ decreases with the dimension of the device and becomes negligible within the temporal resolution of our measurements and the precision of the fits in 3-\textmu m-wide dots. A likely explanation is the following. The incubation time $t_0$ is obtained from the fits under the assumption that the nucleation leads to an appreciable variation of the time-resolved anomalous Hall voltage. If we assume that the volume of the seed domain is independent of the device size (see below), the relative variation of this signal becomes smaller and more difficult to detect in larger devices. In such a case, $t_{\infty}$ represents an artifact of the measurement without any physical meaning. This explanation is supported by the decrease of $t_{\infty}$ with the device diameter $d$ in Fig. \ref{fig:fig4}f. We also note that the rather large value of $t_{\infty}$ cannot be ascribed to any known magnetic dynamics, which should have much shorter timescales in the limit $B_x, V_{\textrm{P}} \rightarrow \infty$. In any case, $t_{\infty}$ does not impact the validity of the model of $t_{0}$, its stochastic properties, and the estimate of the seed volume.

Taking into account this temporal offset, we express the dependence of $t_0$ on $B_x$ and $V_{\textrm{P}}$ as:
\begin{equation}
t_0 (V_{\textrm{P}},B_x) = t_{\infty} + \tau'(V_{\textrm{P}})e^{-B_x/\beta}, \label{eq:expFit_1} 
\end{equation}
where
\begin{gather}
\beta = \frac{(k_{\textrm{B}} T + \eta V_{\textrm{P}}^2)B_x}{\Delta U_B(B_x)} = \frac{(k_{\textrm{B}} T + \eta V_{\textrm{P}}^2)}{cM_{\textrm{s}}}, \label{eq:expFit_2} \\
\tau'(V_{\textrm{P}}) = \tau(V_{\textrm{P}}) e^{\frac{\Delta U_0 - \gamma V_{\textrm{P}}}{k_B T+\eta V_{\textrm{P}}^2}}.
\end{gather}
The parameter $\beta$ represents an effective magnetic field, and $c$ depends on the volume of the seed domain and on the magnetization orientation, as discussed below. Given the dependence of $\beta$ on the temperature and saturation magnetization $M_{\textrm{s}}$, we interpret it as a fictitious field assisting the reversal, analogous to the thermal field used in micromagnetics to account for stochastic thermal effects [\onlinecite{Brown1963,Garcia1998}]. As shown in Fig. \ref{fig:fig4}a, Eq. \ref{eq:expFit_1} captures well the dependence of the nucleation time on the magnetic field. The fits of $t_0$ to Eq. \ref{eq:expFit_1} yield $\beta \approx 23 \pm 2$ mT, with a slight increase of $\beta$ with the voltage, in accordance with Eq. \ref{eq:expFit_2}. Equation \ref{eq:expFit_1} reproduces also the dependence of $t_0$ on the voltage, which is not simply exponential because of Joule heating (Fig. \ref{fig:fig4}b). The good quality of the fits confirms that the nucleation is a dynamical process associated with a single energy barrier and validates our model, which also allows for estimating the volume $\mathcal{V}$ of the reversed domain, as shown below.

\subsection{Estimate of the activation volume}

\begin{figure}
\includegraphics[scale=1]{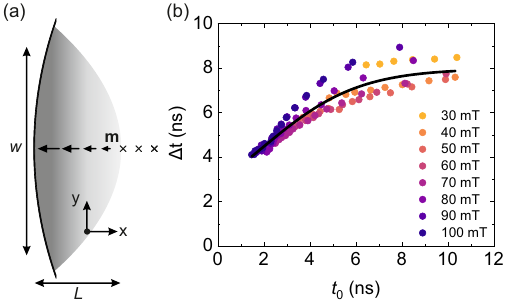}
\caption{(a) Model of the nucleation of the seed domain at the edge of a magnetic dot starting from an initial state magnetized "down". The magnetization $\mathbf{m}$ at the device edge rotates toward the plane. The gray shading represents this rotation from $-z$ (right) to the $xy$ plane (left), which occurs on a lengthscale determined by micromagnetic constraints, i.e., half of the domain wall width $ L = \frac{\delta}{2}$. $w$ is the width of the activated area. (b) Variation of the mean transition time with the mean nucleation time. For a given magnetic field, distinct data points correspond to different pulse amplitudes. The solid line is a fit to Eq. \ref{eq:dtVSt0}.}
\label{fig:fig5}
\end{figure}

We assume that the smallest activated volume has a length $L$ along $x$, starting from the device edge, that equals at least half of a domain wall width $L = \frac{\delta}{2}$ (see Fig. \ref{fig:fig5}a). Here, we estimate $\delta \approx 35$ nm based on the measured magnetic anisotropy and exchange stiffness of about 2 pJ/m reported in the literature for GdFeCo [\onlinecite{Xian-Ying2003,Raasch1994,Radu2018}]. $\mathcal{A} = wt_{\textrm{FM}}$ is the surface of the domain wall in the $yz$ plane, and $t_{\textrm{FM}} = 15$ nm is the thickness of GdFeCo. Before the nucleation, the magnetic moments are oriented at an average angle $\theta$ to the $z$ axis determined by the external field. This angle can be determined from the hysteresis loop in Fig. \ref{fig:fig3}a, for example $\theta =  32^{\circ}$ at 100 mT. With the field angle $\theta _B = 90^{\circ}$, we can write the initial Zeeman energy as $U_B = \frac{\delta}{2} \mathcal{A}M_{\textrm{s}} B\sin\theta$, where $M_{\textrm{s}} =$ 113 kA/m. To calculate the Zeeman energy $U_B'$ after the nucleation, we assume that the domain wall is of Néel type because of the finite DMI of Pt/ferrimagnet bilayers [\onlinecite{Finley2016,Shahbazi2019,Kim2019}] and applied in-plane magnetic field. In this case, $\theta '(x) = 2\arctan(e^{\frac{x}{\delta}})$  [\onlinecite{Mougin2007}] and $U'_B \approx \int_{0}^{\frac{\delta}{2}}\mathcal{A} M_{\textrm{s}} B\sin\theta ' \, dx = 0.48\delta \mathcal{A} M_{\textrm{s}} B$. Thus, $c = \delta \mathcal{A} (0.48 - \frac{\sin\theta}{2})$ and $\beta = \frac{k_{\textrm{B}} T + \eta V_{\textrm{P}}^2}{\delta \mathcal{A} M_{\textrm{s}}(0.48 - \frac{\sin\theta}{2})}$. If we take $k_{\textrm{B}} T + \eta V_{\textrm{P}}^2 \approx k_{\textrm{B}} \cdot 350\, \textrm{K}$, we find $\mathcal{V} = \frac{\delta}{2} \mathcal{A} = 4300$ nm$^3$, $\mathcal{A} = $ 240 nm$^2$, and $w = $ 16 nm. The estimated activation volume is in good agreement with other estimates obtained from measurements of the field-induced magnetization reversal and domain wall depinning, which yield $\mathcal{V} \approx 8\cdot 10^3$ nm$^3$ [\onlinecite{Wernsdorfer1997,Sun2011,Wuth2012}]. This calculation indicates that the activated region is much smaller than the dot size and corresponds to the smallest reversed volume compatible with micromagnetic constraints. We also note that this estimation is independent of the dot dimension as long as the device is much larger than the domain wall width. In future studies, it may be interesting to verify how the device shape influences the size of the seed volume. Estimating the dimension of the seed domain can be useful to identify the critical dimension at which the transition from the domain-wall driven to the macrospin dynamics occurs, which is relevant to the downscaling of devices.

\section[Correlation between $t_0$ and $\Delta t$]{Correlation between $t_0$ and $\Delta t$}

Because the nucleation always precedes the expansion of the seed domain, it is interesting to investigate the relationship between $t_0$ and $\Delta t$. Figure \ref{fig:fig5}b shows that $t_0$ and $\Delta t$ correlate over several combinations of the pulse amplitude and magnetic field. In general, we find that the data points follow the same universal curve, and the longer $t_0$, the shorter $\Delta t$ is. The relation is not simply linear. However, a simple way to explain the overall trend in Fig. \ref{fig:fig5}b is based on the assumption that temperature mediates the correlation between the two times. The domain wall speed $v$ in a ferrimagnet depends on several parameters but, if we retain only those that are in principle temperature dependent, we may write $v \sim \frac{\delta\theta_{\textrm{SHE}}}{\alpha(g_2M_1 + g_1M_2)}$ ($g_i$ and $M_i$ are the Landé g-factor and the saturation magnetization of the $i^{\textrm{th}}$ sublattice) [\onlinecite{Martinez2019}]. The spin Hall angle $\theta_{\textrm{SHE}}$ is not expected to vary significantly with temperature since the spin Hall effect in Pt is of intrinsic origin, nor is the effective damping $\alpha$ of GdFeCo  [\onlinecite{Kim2019}]. Since the exchange stiffness and anisotropy can be expressed as powers of $\sim \left(\frac{M(T)}{M_{\textrm{s}}}\right)^p$ of the net magnetization [\onlinecite{Grimaldi2020,Moreno2016,Hirata2018}], which in turn depends on temperature according to a power law of the form $M(T) \sim \left(1 - \frac{T}{T_{\textrm{c}}}\right)^q$, we can write that $\Delta t \sim 1/v \sim \left(g_2M_1 + g_1M_2\right)\sqrt{\frac{k_{\textrm{eff}}}{A_{\textrm{ex}}}} \sim \left(1 - \frac{T}{T_{\textrm{c}}}\right)^{k}$, with $T_{\textrm{c}}$ the Curie temperature and $k$ an appropriate exponential. Further, finite-element COMSOL simulations show that the increase in time of temperature caused by Joule heating is governed by a law of the type $T(t) = T_0 + \Delta T\left(1 - e^{-\frac{t}{\tau_T}}\right)$, where $T_0$ is the ambient temperature, $\Delta T$ the maximum temperature variation, and $\tau_T$ the corresponding time constant. Upon setting $t = t_0$, we obtain
\begin{align}
\Delta t \sim \left(E + Fe^{-\frac{t_0}{\tau_T}}\right)^k,
\label{eq:dtVSt0}
\end{align}
where $E^k \approx 8$ ns and $(E+F)^k \approx 2.8$ ns are the transition time in the long and short nucleation time regime, respectively. As shown in Fig. \ref{fig:fig5}b, Eq. \ref{eq:dtVSt0} can fit well the dependence of $\Delta t$ on $t_0$ with $k \approx -0.6$ and $\tau_T \approx 2$ ns despite our simplifications. This result supports our hypothesis of a correlation between the two times mediated by temperature.

\section[Statistical distributions of $t_0$ and $\Delta t$]{Statistical distributions of $t_0$ and $\Delta t$}

\begin{figure}
\includegraphics[scale=1]{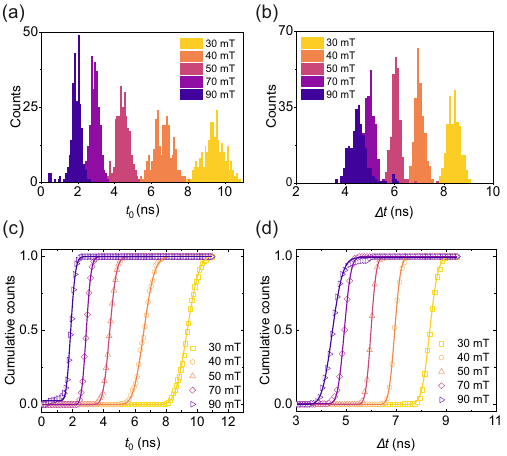} 
\caption{(a), (b) Distributions of the nucleation and transition time, respectively, at different magnetic fields and for a pulse amplitude $V_{\textrm{P}}=30$ V. The datasets include 250 measurements. The bin size is 100 ps, larger than the nominal temporal resolution of 50 ps, for ease of visualization. (c), (d) Cumulative distributions obtained from the data in (a)-(b) and fits (solid lines) to the log-normal and Gamma cumulative distribution function, respectively.}
\label{fig:fig6}
\end{figure}

Time-resolved Hall measurements provide insight also into the stochastic nature of the dynamics. Figure \ref{fig:fig6}a,b shows exemplary distributions of the nucleation and transition times at different in-plane magnetic fields. As the field increases, both the mean and standard deviation of the distributions decrease [\onlinecite{Grimaldi2020,Krizakova2020,Hahn2016}], in accordance with the trends in Fig. \ref{fig:fig4}a,c. The distributions show a similar trend when the control parameter is the pulse amplitude because in the limit of large $V_{\textrm{P}}$ the dynamics is essentially dictated by SOT [\onlinecite{Vincent2015}].

As a first attempt, we may choose the Gaussian function to describe the variability of $t_0$ and $\Delta t$ because thermal fluctuations are usually assumed to be Gaussian-distributed [\onlinecite{Brown1963,Martinez2007}]. We find, indeed, that the Gaussian function fits reasonably well the distributions of both $t_0$ and $\Delta  t$ (not shown). However, we argue that this function does not provide the correct statistics of the two characteristic times. In fact, both $t_0$ and $\Delta t$ are lower bounded, i.e., they cannot be smaller than 0. Therefore, we expect that in the limit of large $V_{\textrm{P}}$ and $B_x$ the skewness of the distributions increases because of the accumulation of data points near 0. Although this tendency is not clearly observable in Fig. \ref{fig:fig6}a,b, skewed distributions were found in former measurements of similar ferro- and ferrimagnetic samples [\onlinecite{Bouquin2021,Grimaldi2020,Krizakova2020,Sala2021,Tomita2008,Zhao2012}]. This evidence indicates the need for alternative statistical models, which, in the case of $\Delta t$, cannot simply be the exponential function that is sometimes used to model the depinning time of domain walls [\onlinecite{Cui2010,Bultynck2018,Burrowes2010}]. Here we adopt a phenomenological approach and propose distinct distributions for $t_0$ and $\Delta t$ with a direct physical interpretation. 

\subsection{Interpretation of the stochasticity of $t_0$}

According to Eq. \ref{eq:basic_t0}, the nucleation time is a function of the energy barrier $\Delta U$, hence, its variability stems directly from fluctuations of $\Delta U$. If we assume that such fluctuations are of thermal origin and hence Gaussian-distributed [\onlinecite{Brown1963,Martinez2007}], then the statistics of $t_0$ is log-normal [\onlinecite{Taniguchi2014}] because the exponential function $Y = e^X$ of a normally-distributed random variable $X$ has log-normal distribution. As shown in Fig. \ref{fig:fig6}c, the cumulative distribution of $t_0$ is well fitted by the cumulative log-normal distribution function. Although the Gaussian and gamma functions fit equally well to the cumulative distributions in Fig. \ref{fig:fig6}c, we consider the log-normal function a mathematically- and physically-sound statistics for the nucleation time. It is associated with the overcome of an energy barrier, independently of whether SOT or another force drives the nucleation. 

Indeed, in a mean-field model, the energy barrier $\Delta U$ is a macroscopic quantity obtained by summing the microscopic energies $U_i$ of all the magnetic moments involved in the nucleation of a seed domain. The macroscopic fluctuations of $\Delta U$ are approximately the sum of the microscopic random variations of $U_i$, that is, $\Delta U$ is a random variable with Gaussian distribution. As a consequence, $t_0 \sim e^{\frac{\Delta U} {k_{\textrm{B}}T} } \approx \prod_i e^{\frac{U_i} {k_{\textrm{B}}T} }$ is a log-normal variable because the product of many independent random variables approaches the log-normal distribution, in the same way as the sum of many independent variables is normally distributed. This reasoning indicates that, at the macroscopic level, independent thermal fluctuations are multiplicative, not additive. Moreover, it implies that the standard deviation of $t_0$ grows exponentially with the number of involved magnetic moments, that is, with the dimension of the seed domain. This implication may be examined in future works. Note that a log-normal distribution only admits positive values, consistently with $t_0, \Delta t > 0$.


\subsection{Interpretation of the stochasticity of $\Delta t$}

The motion of a domain wall is influenced by both thermal fluctuations and local variations of the pinning potential. Micromagnetic simulations that only include the thermal field predict a Gaussian distribution of depinning times [\onlinecite{Martinez2007}]. However, previous experimental work has found evidence of more complex statistics [\onlinecite{Im2009,Vincent2015}], and exponential [\onlinecite{Burrowes2010,Bedau2010}], log-normal [\onlinecite{Taniguchi2014}], and gamma [\onlinecite{Metaxas2013}] functions have been used to describe the distributions of depinning times from notches or constrictions. Simple physical considerations, however, support the choice of a specific distribution. The transition time is associated with the high-speed and nearly-continuous motion of the domain wall in a small and confined device [\onlinecite{Baumgartner2017,Sala2022}]. During its propagation, the domain wall can encounter multiple independent pinning sites in the form of local variations of the magnetic properties, surface and edge roughness, and device geometry. The transition time is therefore the sum of the times $\Delta t_i$ taken to move from one pinning point to the next and the depinning times at each site $t_j$: $\Delta t = \sum_i \Delta t_i + \sum_j t_j$. We assume for simplicity that the time intervals $\Delta t_i$ are nearly deterministic and determined uniquely by the domain wall speed, which is in turn set by the external magnetic field and SOT. In contrast, we consider the depinning times $t_j$ as the main source of stochasticity. We show now that the nature of the pinning determines the overall statistical distribution of $\Delta t$. 

If the depinning process is statistically described by an exponential distribution, as often assumed [\onlinecite{Attane2006,Burrowes2010,Bultynck2018,Cui2010}], then the sum $\sum_j t_j$ and, hence, $\Delta t$ is mathematically a random variable with gamma distribution [\onlinecite{Gupta1960,mcGill1965}]. If, instead, we assume that each depinning time behaves similarly to $t_0$ and is thus a log-normal variable, then $\Delta t$ is a random variable with no explicit statistical distribution because the sum of log-normal variables resembles but is not a log-normal distribution. In practice, the overall pinning landscape may encompass different pinning potentials with distinct distribution functions, such that the overall statistic of $\Delta t$ depends on the details of the sample [\onlinecite{Im2009}]. In the present case, we find that the cumulative gamma function fits well to the cumulative distributions of $\Delta t$, as shown in Fig. \ref{fig:fig6}d. A more extensive investigation as a function of the sample material and geometry is needed to ascertain the actual nature of the stochasticity of $\Delta t$. We note, however, that the gamma distribution offers an interesting connection with other physical phenomena. Gamma-distributed events are particular cases of a Lévy process, which is the simplest process consisting of a continuous motion with jumps of random size occurring at random times [\onlinecite{Applebaum2004}]. For instance, the Poisson process and the Brownian motion, which is the continuous-time version of a random walk, are Lévy processes. The field- or current-driven dynamics of domain walls share these features because the domain expansion may be interpreted as the continuous motion of a domain wall interspersed by random occurrences (pinning) that alter its speed. The accumulation of successive random events represents a fundamental difference between the variability of $\Delta t$ and $t_0$.

Our considerations on the stochasticity of $t_0$ and $\Delta t$ are based on the physical processes described by the nucleation and transition times and do not refer to specific experimental conditions. They are general and remain valid independently of the amplitude and duration of the electric pulses or the strength of the magnetic field. Moreover, our model is independent of the magnetic properties of the sample under investigation and is expected to lose validity only when the magnetic dynamics approach the macrospin regime, i.e., in devices smaller than a few tens of nm.

\section{Conclusions}

We have analyzed the dependence of the nucleation time $t_0$ and transition time $\Delta t$ on the pulse amplitude $V_{\textrm{P}}$ and assisting in-plane magnetic field $B_x$ during the SOT-induced switching of perpendicularly-magnetized GdFeCo dots. 
We have found that the decrease of $\Delta t$ with $V_{\textrm{P}}$ and $B_x$ can be interpreted according to existing models of the domain wall motion driven by SOT. On the other hand, we have proposed a new framework to understand the variation of $t_0$ with current and field and provided an estimate of the dimensions of the nucleated domain. In our model, $t_0$ depends on the activation energy required to nucleate a seed domain against a single energy barrier determined by the magnetic anisotropy, the Zeeman energy, and the DMI. We have also revealed the existence of a correlation between $t_0$ and $\Delta t$ established by the temperature variation during the switching.

Finally, we have discussed the physical origin of the stochasticity of the two times, which decreases with strong pulses or magnetic fields. The variability of $t_0$ is associated with thermal fluctuations of a single energy barrier and is captured by a log-normal function. Instead, the distributions of $\Delta t$ can be interpreted in terms of a cumulative process of random events associated with distinct pinning sites. The exact analytical function describing the statistical distribution of $\Delta t$ depends on the details of the pinning landscape, but can be approximated by a gamma distribution. We expect these statistical considerations to be valid also for the STT-induced magnetization switching because the reversal mechanism (nucleation and domain expansion) is to some extents similar to that triggered by SOT, and we do not make explicit reference to the driving force of the dynamics. Overall, our study provides new insight into the deterministic and stochastic aspects of the magnetization switching induced by electric currents.

\section{Acknowledgement}

We acknowledge the support of the Swiss National Science Foundation (Grant No. 200020-200465).

\section{Bibliography}

\bibliography{library.bib}

\begin{thebibliography}{84}%
\makeatletter
\providecommand \@ifxundefined [1]{%
 \@ifx{#1\undefined}
}%
\providecommand \@ifnum [1]{%
 \ifnum #1\expandafter \@firstoftwo
 \else \expandafter \@secondoftwo
 \fi
}%
\providecommand \@ifx [1]{%
 \ifx #1\expandafter \@firstoftwo
 \else \expandafter \@secondoftwo
 \fi
}%
\providecommand \natexlab [1]{#1}%
\providecommand \enquote  [1]{``#1''}%
\providecommand \bibnamefont  [1]{#1}%
\providecommand \bibfnamefont [1]{#1}%
\providecommand \citenamefont [1]{#1}%
\providecommand \href@noop [0]{\@secondoftwo}%
\providecommand \href [0]{\begingroup \@sanitize@url \@href}%
\providecommand \@href[1]{\@@startlink{#1}\@@href}%
\providecommand \@@href[1]{\endgroup#1\@@endlink}%
\providecommand \@sanitize@url [0]{\catcode `\\12\catcode `\$12\catcode
  `\&12\catcode `\#12\catcode `\^12\catcode `\_12\catcode `\%12\relax}%
\providecommand \@@startlink[1]{}%
\providecommand \@@endlink[0]{}%
\providecommand \url  [0]{\begingroup\@sanitize@url \@url }%
\providecommand \@url [1]{\endgroup\@href {#1}{\urlprefix }}%
\providecommand \urlprefix  [0]{URL }%
\providecommand \Eprint [0]{\href }%
\providecommand \doibase [0]{http://dx.doi.org/}%
\providecommand \selectlanguage [0]{\@gobble}%
\providecommand \bibinfo  [0]{\@secondoftwo}%
\providecommand \bibfield  [0]{\@secondoftwo}%
\providecommand \translation [1]{[#1]}%
\providecommand \BibitemOpen [0]{}%
\providecommand \bibitemStop [0]{}%
\providecommand \bibitemNoStop [0]{.\EOS\space}%
\providecommand \EOS [0]{\spacefactor3000\relax}%
\providecommand \BibitemShut  [1]{\csname bibitem#1\endcsname}%
\let\auto@bib@innerbib\@empty
\bibitem [{\citenamefont {Myers}\ \emph {et~al.}(1999)\citenamefont {Myers},
  \citenamefont {Ralph}, \citenamefont {Katine}, \citenamefont {Louie},\ and\
  \citenamefont {Buhrman}}]{Myers1999}%
  \BibitemOpen
  \bibfield  {author} {\bibinfo {author} {\bibfnamefont {E.~B.}\ \bibnamefont
  {Myers}}, \bibinfo {author} {\bibfnamefont {D.~C.}\ \bibnamefont {Ralph}},
  \bibinfo {author} {\bibfnamefont {J.~A.}\ \bibnamefont {Katine}}, \bibinfo
  {author} {\bibfnamefont {R.~N.}\ \bibnamefont {Louie}}, \ and\ \bibinfo
  {author} {\bibfnamefont {R.~A.}\ \bibnamefont {Buhrman}},\ }\bibfield
  {title} {\enquote {\bibinfo {title} {{Current-Induced Switching of Domains in
  Magnetic Multilayer Devices}},}\ }\href {\doibase
  10.1126/science.285.5429.867} {\bibfield  {journal} {\bibinfo  {journal}
  {Science}\ }\textbf {\bibinfo {volume} {285}},\ \bibinfo {pages} {867--870}
  (\bibinfo {year} {1999})}\BibitemShut {NoStop}%
\bibitem [{\citenamefont {Wegrowe}\ \emph {et~al.}(1999)\citenamefont
  {Wegrowe}, \citenamefont {Fruchart}, \citenamefont {Nozi{\`{e}}res},
  \citenamefont {Givord}, \citenamefont {Rousseaux}, \citenamefont {Decanini},\
  and\ \citenamefont {Ansermet}}]{Wegrowe1999}%
  \BibitemOpen
  \bibfield  {author} {\bibinfo {author} {\bibfnamefont {J.-E.}\ \bibnamefont
  {Wegrowe}}, \bibinfo {author} {\bibfnamefont {O.}~\bibnamefont {Fruchart}},
  \bibinfo {author} {\bibfnamefont {J.-P.}\ \bibnamefont {Nozi{\`{e}}res}},
  \bibinfo {author} {\bibfnamefont {D.}~\bibnamefont {Givord}}, \bibinfo
  {author} {\bibfnamefont {F.}~\bibnamefont {Rousseaux}}, \bibinfo {author}
  {\bibfnamefont {D.}~\bibnamefont {Decanini}}, \ and\ \bibinfo {author}
  {\bibfnamefont {J.~P.}\ \bibnamefont {Ansermet}},\ }\bibfield  {title}
  {\enquote {\bibinfo {title} {{Arrays of ultrathin monocrystalline
  submicrometer-sized Fe dots: N{\'{e}}el–Brown relaxation and activation
  volume}},}\ }\href {\doibase 10.1063/1.370842} {\bibfield  {journal}
  {\bibinfo  {journal} {Journal of Applied Physics}\ }\textbf {\bibinfo
  {volume} {86}},\ \bibinfo {pages} {1028--1034} (\bibinfo {year}
  {1999})}\BibitemShut {NoStop}%
\bibitem [{\citenamefont {Stiles}\ and\ \citenamefont
  {Zangwill}(2002)}]{Stiles2002}%
  \BibitemOpen
  \bibfield  {author} {\bibinfo {author} {\bibfnamefont {M.~D.}\ \bibnamefont
  {Stiles}}\ and\ \bibinfo {author} {\bibfnamefont {A.}~\bibnamefont
  {Zangwill}},\ }\bibfield  {title} {\enquote {\bibinfo {title} {{Anatomy of
  spin-transfer torque}},}\ }\href {\doibase 10.1103/PhysRevB.66.014407}
  {\bibfield  {journal} {\bibinfo  {journal} {Physical Review B}\ }\textbf
  {\bibinfo {volume} {66}},\ \bibinfo {pages} {014407} (\bibinfo {year}
  {2002})}\BibitemShut {NoStop}%
\bibitem [{\citenamefont {Miron}\ \emph {et~al.}(2011)\citenamefont {Miron},
  \citenamefont {Garello}, \citenamefont {Gaudin}, \citenamefont {Zermatten},
  \citenamefont {Costache}, \citenamefont {Auffret}, \citenamefont {Bandiera},
  \citenamefont {Rodmacq}, \citenamefont {Schuhl},\ and\ \citenamefont
  {Gambardella}}]{Miron2011}%
  \BibitemOpen
  \bibfield  {author} {\bibinfo {author} {\bibfnamefont {I.~M.}\ \bibnamefont
  {Miron}}, \bibinfo {author} {\bibfnamefont {K.}~\bibnamefont {Garello}},
  \bibinfo {author} {\bibfnamefont {G.}~\bibnamefont {Gaudin}}, \bibinfo
  {author} {\bibfnamefont {P.-J.}\ \bibnamefont {Zermatten}}, \bibinfo {author}
  {\bibfnamefont {M.~V.}\ \bibnamefont {Costache}}, \bibinfo {author}
  {\bibfnamefont {S.}~\bibnamefont {Auffret}}, \bibinfo {author} {\bibfnamefont
  {S.}~\bibnamefont {Bandiera}}, \bibinfo {author} {\bibfnamefont
  {B.}~\bibnamefont {Rodmacq}}, \bibinfo {author} {\bibfnamefont
  {A.}~\bibnamefont {Schuhl}}, \ and\ \bibinfo {author} {\bibfnamefont
  {P.}~\bibnamefont {Gambardella}},\ }\bibfield  {title} {\enquote {\bibinfo
  {title} {{Perpendicular switching of a single ferromagnetic layer induced by
  in-plane current injection}},}\ }\href {\doibase 10.1038/nature10309}
  {\bibfield  {journal} {\bibinfo  {journal} {Nature}\ }\textbf {\bibinfo
  {volume} {476}},\ \bibinfo {pages} {189--193} (\bibinfo {year}
  {2011})}\BibitemShut {NoStop}%
\bibitem [{\citenamefont {Liu}\ \emph {et~al.}(2012)\citenamefont {Liu},
  \citenamefont {Pai}, \citenamefont {Li}, \citenamefont {Tseng}, \citenamefont
  {Ralph},\ and\ \citenamefont {Buhrman}}]{Liu2012a}%
  \BibitemOpen
  \bibfield  {author} {\bibinfo {author} {\bibfnamefont {L.}~\bibnamefont
  {Liu}}, \bibinfo {author} {\bibfnamefont {C.-F.}\ \bibnamefont {Pai}},
  \bibinfo {author} {\bibfnamefont {Y.}~\bibnamefont {Li}}, \bibinfo {author}
  {\bibfnamefont {H.~W.}\ \bibnamefont {Tseng}}, \bibinfo {author}
  {\bibfnamefont {D.~C.}\ \bibnamefont {Ralph}}, \ and\ \bibinfo {author}
  {\bibfnamefont {R.~A.}\ \bibnamefont {Buhrman}},\ }\bibfield  {title}
  {\enquote {\bibinfo {title} {{Spin-Torque Switching with the Giant Spin Hall
  Effect of Tantalum}},}\ }\href {\doibase 10.1126/science.1218197} {\bibfield
  {journal} {\bibinfo  {journal} {Science}\ }\textbf {\bibinfo {volume}
  {336}},\ \bibinfo {pages} {555--558} (\bibinfo {year} {2012})},\ \Eprint
  {http://arxiv.org/abs/1203.2875} {arXiv:1203.2875} \BibitemShut {NoStop}%
\bibitem [{\citenamefont {Manchon}\ \emph {et~al.}(2019)\citenamefont
  {Manchon}, \citenamefont {{\v{Z}}elezn{\'{y}}}, \citenamefont {Miron},
  \citenamefont {Jungwirth}, \citenamefont {Sinova}, \citenamefont {Thiaville},
  \citenamefont {Garello},\ and\ \citenamefont {Gambardella}}]{Manchon2019}%
  \BibitemOpen
  \bibfield  {author} {\bibinfo {author} {\bibfnamefont {A.}~\bibnamefont
  {Manchon}}, \bibinfo {author} {\bibfnamefont {J.}~\bibnamefont
  {{\v{Z}}elezn{\'{y}}}}, \bibinfo {author} {\bibfnamefont {I.~M.}\
  \bibnamefont {Miron}}, \bibinfo {author} {\bibfnamefont {T.}~\bibnamefont
  {Jungwirth}}, \bibinfo {author} {\bibfnamefont {J.}~\bibnamefont {Sinova}},
  \bibinfo {author} {\bibfnamefont {A.}~\bibnamefont {Thiaville}}, \bibinfo
  {author} {\bibfnamefont {K.}~\bibnamefont {Garello}}, \ and\ \bibinfo
  {author} {\bibfnamefont {P.}~\bibnamefont {Gambardella}},\ }\bibfield
  {title} {\enquote {\bibinfo {title} {{Current-induced spin-orbit torques in
  ferromagnetic and antiferromagnetic systems}},}\ }\href {\doibase
  10.1103/RevModPhys.91.035004} {\bibfield  {journal} {\bibinfo  {journal}
  {Reviews of Modern Physics}\ }\textbf {\bibinfo {volume} {91}} (\bibinfo
  {year} {2019}),\ 10.1103/RevModPhys.91.035004},\ \Eprint
  {http://arxiv.org/abs/1801.09636} {arXiv:1801.09636} \BibitemShut {NoStop}%
\bibitem [{\citenamefont {Khvalkovskiy}\ \emph {et~al.}(2013)\citenamefont
  {Khvalkovskiy}, \citenamefont {Apalkov}, \citenamefont {Watts}, \citenamefont
  {Chepulskii}, \citenamefont {Beach}, \citenamefont {Ong}, \citenamefont
  {Tang}, \citenamefont {Driskill-Smith}, \citenamefont {Butler}, \citenamefont
  {Visscher}, \citenamefont {Lottis}, \citenamefont {Chen}, \citenamefont
  {Nikitin},\ and\ \citenamefont {Krounbi}}]{Khvalkovskiy2013a}%
  \BibitemOpen
  \bibfield  {author} {\bibinfo {author} {\bibfnamefont {A.~V.}\ \bibnamefont
  {Khvalkovskiy}}, \bibinfo {author} {\bibfnamefont {D.}~\bibnamefont
  {Apalkov}}, \bibinfo {author} {\bibfnamefont {S.}~\bibnamefont {Watts}},
  \bibinfo {author} {\bibfnamefont {R.}~\bibnamefont {Chepulskii}}, \bibinfo
  {author} {\bibfnamefont {R.~S.}\ \bibnamefont {Beach}}, \bibinfo {author}
  {\bibfnamefont {A.}~\bibnamefont {Ong}}, \bibinfo {author} {\bibfnamefont
  {X.}~\bibnamefont {Tang}}, \bibinfo {author} {\bibfnamefont {A.}~\bibnamefont
  {Driskill-Smith}}, \bibinfo {author} {\bibfnamefont {W.~H.}\ \bibnamefont
  {Butler}}, \bibinfo {author} {\bibfnamefont {P.~B.}\ \bibnamefont
  {Visscher}}, \bibinfo {author} {\bibfnamefont {D.}~\bibnamefont {Lottis}},
  \bibinfo {author} {\bibfnamefont {E.}~\bibnamefont {Chen}}, \bibinfo {author}
  {\bibfnamefont {V.}~\bibnamefont {Nikitin}}, \ and\ \bibinfo {author}
  {\bibfnamefont {M.}~\bibnamefont {Krounbi}},\ }\bibfield  {title} {\enquote
  {\bibinfo {title} {{Basic principles of STT-MRAM cell operation in memory
  arrays}},}\ }\href {\doibase 10.1088/0022-3727/46/13/139601} {\bibfield
  {journal} {\bibinfo  {journal} {Journal of Physics D: Applied Physics}\
  }\textbf {\bibinfo {volume} {46}},\ \bibinfo {pages} {139601} (\bibinfo
  {year} {2013})}\BibitemShut {NoStop}%
\bibitem [{\citenamefont {Krizakova}\ \emph {et~al.}(2022)\citenamefont
  {Krizakova}, \citenamefont {Perumkunnil}, \citenamefont {Couet},
  \citenamefont {Gambardella},\ and\ \citenamefont {Garello}}]{Krizakova2022}%
  \BibitemOpen
  \bibfield  {author} {\bibinfo {author} {\bibfnamefont {V.}~\bibnamefont
  {Krizakova}}, \bibinfo {author} {\bibfnamefont {M.}~\bibnamefont
  {Perumkunnil}}, \bibinfo {author} {\bibfnamefont {S.}~\bibnamefont {Couet}},
  \bibinfo {author} {\bibfnamefont {P.}~\bibnamefont {Gambardella}}, \ and\
  \bibinfo {author} {\bibfnamefont {K.}~\bibnamefont {Garello}},\ }\bibfield
  {title} {\enquote {\bibinfo {title} {{Spin-orbit torque switching of magnetic
  tunnel junctions for memory applications}},}\ }\href {\doibase
  10.1016/j.jmmm.2022.169692} {\bibfield  {journal} {\bibinfo  {journal}
  {Journal of Magnetism and Magnetic Materials}\ }\textbf {\bibinfo {volume}
  {562}},\ \bibinfo {pages} {169692} (\bibinfo {year} {2022})}\BibitemShut
  {NoStop}%
\bibitem [{\citenamefont {Sun}(2000)}]{Sun2000}%
  \BibitemOpen
  \bibfield  {author} {\bibinfo {author} {\bibfnamefont {J.~Z.}\ \bibnamefont
  {Sun}},\ }\bibfield  {title} {\enquote {\bibinfo {title} {{Spin-current
  interaction with a monodomain magnetic body: A model study}},}\ }\href
  {\doibase 10.1103/PhysRevB.62.570} {\bibfield  {journal} {\bibinfo  {journal}
  {Physical Review B}\ }\textbf {\bibinfo {volume} {62}},\ \bibinfo {pages}
  {570--578} (\bibinfo {year} {2000})}\BibitemShut {NoStop}%
\bibitem [{\citenamefont {Lee}\ \emph {et~al.}(2013)\citenamefont {Lee},
  \citenamefont {Lee}, \citenamefont {Min},\ and\ \citenamefont
  {Lee}}]{Lee2013}%
  \BibitemOpen
  \bibfield  {author} {\bibinfo {author} {\bibfnamefont {K.-S.}\ \bibnamefont
  {Lee}}, \bibinfo {author} {\bibfnamefont {S.-W.}\ \bibnamefont {Lee}},
  \bibinfo {author} {\bibfnamefont {B.-C.}\ \bibnamefont {Min}}, \ and\
  \bibinfo {author} {\bibfnamefont {K.-J.}\ \bibnamefont {Lee}},\ }\bibfield
  {title} {\enquote {\bibinfo {title} {{Threshold current for switching of a
  perpendicular magnetic layer induced by spin Hall effect}},}\ }\href
  {\doibase 10.1063/1.4798288} {\bibfield  {journal} {\bibinfo  {journal}
  {Applied Physics Letters}\ }\textbf {\bibinfo {volume} {102}},\ \bibinfo
  {pages} {112410} (\bibinfo {year} {2013})}\BibitemShut {NoStop}%
\bibitem [{\citenamefont {Bernstein}\ \emph {et~al.}(2011)\citenamefont
  {Bernstein}, \citenamefont {Br{\"{a}}uer}, \citenamefont {Kukreja},
  \citenamefont {St{\"{o}}hr}, \citenamefont {Hauet}, \citenamefont
  {Cucchiara}, \citenamefont {Mangin}, \citenamefont {Katine}, \citenamefont
  {Tyliszczak}, \citenamefont {Chou},\ and\ \citenamefont
  {Acremann}}]{Bernstein2011}%
  \BibitemOpen
  \bibfield  {author} {\bibinfo {author} {\bibfnamefont {D.~P.}\ \bibnamefont
  {Bernstein}}, \bibinfo {author} {\bibfnamefont {B.}~\bibnamefont
  {Br{\"{a}}uer}}, \bibinfo {author} {\bibfnamefont {R.}~\bibnamefont
  {Kukreja}}, \bibinfo {author} {\bibfnamefont {J.}~\bibnamefont
  {St{\"{o}}hr}}, \bibinfo {author} {\bibfnamefont {T.}~\bibnamefont {Hauet}},
  \bibinfo {author} {\bibfnamefont {J.}~\bibnamefont {Cucchiara}}, \bibinfo
  {author} {\bibfnamefont {S.}~\bibnamefont {Mangin}}, \bibinfo {author}
  {\bibfnamefont {J.~A.}\ \bibnamefont {Katine}}, \bibinfo {author}
  {\bibfnamefont {T.}~\bibnamefont {Tyliszczak}}, \bibinfo {author}
  {\bibfnamefont {K.~W.}\ \bibnamefont {Chou}}, \ and\ \bibinfo {author}
  {\bibfnamefont {Y.}~\bibnamefont {Acremann}},\ }\bibfield  {title} {\enquote
  {\bibinfo {title} {{Nonuniform switching of the perpendicular magnetization
  in a spin-torque-driven magnetic nanopillar}},}\ }\href {\doibase
  10.1103/PhysRevB.83.180410} {\bibfield  {journal} {\bibinfo  {journal}
  {Physical Review B}\ }\textbf {\bibinfo {volume} {83}},\ \bibinfo {pages}
  {180410} (\bibinfo {year} {2011})}\BibitemShut {NoStop}%
\bibitem [{\citenamefont {Baumgartner}\ \emph {et~al.}(2017)\citenamefont
  {Baumgartner}, \citenamefont {Garello}, \citenamefont {Mendil}, \citenamefont
  {Avci}, \citenamefont {Grimaldi}, \citenamefont {Murer}, \citenamefont
  {Feng}, \citenamefont {Gabureac}, \citenamefont {Stamm}, \citenamefont
  {Acremann}, \citenamefont {Finizio}, \citenamefont {Wintz}, \citenamefont
  {Raabe},\ and\ \citenamefont {Gambardella}}]{Baumgartner2017}%
  \BibitemOpen
  \bibfield  {author} {\bibinfo {author} {\bibfnamefont {M.}~\bibnamefont
  {Baumgartner}}, \bibinfo {author} {\bibfnamefont {K.}~\bibnamefont
  {Garello}}, \bibinfo {author} {\bibfnamefont {J.}~\bibnamefont {Mendil}},
  \bibinfo {author} {\bibfnamefont {C.~O.}\ \bibnamefont {Avci}}, \bibinfo
  {author} {\bibfnamefont {E.}~\bibnamefont {Grimaldi}}, \bibinfo {author}
  {\bibfnamefont {C.}~\bibnamefont {Murer}}, \bibinfo {author} {\bibfnamefont
  {J.}~\bibnamefont {Feng}}, \bibinfo {author} {\bibfnamefont {M.}~\bibnamefont
  {Gabureac}}, \bibinfo {author} {\bibfnamefont {C.}~\bibnamefont {Stamm}},
  \bibinfo {author} {\bibfnamefont {Y.}~\bibnamefont {Acremann}}, \bibinfo
  {author} {\bibfnamefont {S.}~\bibnamefont {Finizio}}, \bibinfo {author}
  {\bibfnamefont {S.}~\bibnamefont {Wintz}}, \bibinfo {author} {\bibfnamefont
  {J.}~\bibnamefont {Raabe}}, \ and\ \bibinfo {author} {\bibfnamefont
  {P.}~\bibnamefont {Gambardella}},\ }\bibfield  {title} {\enquote {\bibinfo
  {title} {{Spatially and time-resolved magnetization dynamics driven by
  spin-orbit torques}},}\ }\href {\doibase 10.1038/nnano.2017.151} {\bibfield
  {journal} {\bibinfo  {journal} {Nature Nanotechnology}\ }\textbf {\bibinfo
  {volume} {12}},\ \bibinfo {pages} {980--986} (\bibinfo {year} {2017})},\
  \Eprint {http://arxiv.org/abs/arXiv:1704.06402v1} {arXiv:arXiv:1704.06402v1}
  \BibitemShut {NoStop}%
\bibitem [{\citenamefont {Sun}\ \emph {et~al.}(2011)\citenamefont {Sun},
  \citenamefont {Robertazzi}, \citenamefont {Nowak}, \citenamefont
  {Trouilloud}, \citenamefont {Hu}, \citenamefont {Abraham}, \citenamefont
  {Gaidis}, \citenamefont {Brown}, \citenamefont {O'Sullivan}, \citenamefont
  {Gallagher},\ and\ \citenamefont {Worledge}}]{Sun2011}%
  \BibitemOpen
  \bibfield  {author} {\bibinfo {author} {\bibfnamefont {J.~Z.}\ \bibnamefont
  {Sun}}, \bibinfo {author} {\bibfnamefont {R.~P.}\ \bibnamefont {Robertazzi}},
  \bibinfo {author} {\bibfnamefont {J.}~\bibnamefont {Nowak}}, \bibinfo
  {author} {\bibfnamefont {P.~L.}\ \bibnamefont {Trouilloud}}, \bibinfo
  {author} {\bibfnamefont {G.}~\bibnamefont {Hu}}, \bibinfo {author}
  {\bibfnamefont {D.~W.}\ \bibnamefont {Abraham}}, \bibinfo {author}
  {\bibfnamefont {M.~C.}\ \bibnamefont {Gaidis}}, \bibinfo {author}
  {\bibfnamefont {S.~L.}\ \bibnamefont {Brown}}, \bibinfo {author}
  {\bibfnamefont {E.~J.}\ \bibnamefont {O'Sullivan}}, \bibinfo {author}
  {\bibfnamefont {W.~J.}\ \bibnamefont {Gallagher}}, \ and\ \bibinfo {author}
  {\bibfnamefont {D.~C.}\ \bibnamefont {Worledge}},\ }\bibfield  {title}
  {\enquote {\bibinfo {title} {{Effect of subvolume excitation and spin-torque
  efficiency on magnetic switching}},}\ }\href {\doibase
  10.1103/PhysRevB.84.064413} {\bibfield  {journal} {\bibinfo  {journal}
  {Physical Review B}\ }\textbf {\bibinfo {volume} {84}},\ \bibinfo {pages}
  {064413} (\bibinfo {year} {2011})}\BibitemShut {NoStop}%
\bibitem [{\citenamefont {Chaves-O'Flynn}\ \emph {et~al.}(2015)\citenamefont
  {Chaves-O'Flynn}, \citenamefont {Wolf}, \citenamefont {Sun},\ and\
  \citenamefont {Kent}}]{Chaves-OFlynn2015}%
  \BibitemOpen
  \bibfield  {author} {\bibinfo {author} {\bibfnamefont {G.~D.}\ \bibnamefont
  {Chaves-O'Flynn}}, \bibinfo {author} {\bibfnamefont {G.}~\bibnamefont
  {Wolf}}, \bibinfo {author} {\bibfnamefont {J.~Z.}\ \bibnamefont {Sun}}, \
  and\ \bibinfo {author} {\bibfnamefont {A.~D.}\ \bibnamefont {Kent}},\
  }\bibfield  {title} {\enquote {\bibinfo {title} {{Thermal Stability of
  Magnetic States in Circular Thin-Film Nanomagnets with Large Perpendicular
  Magnetic Anisotropy}},}\ }\href {\doibase 10.1103/PhysRevApplied.4.024010}
  {\bibfield  {journal} {\bibinfo  {journal} {Physical Review Applied}\
  }\textbf {\bibinfo {volume} {4}},\ \bibinfo {pages} {024010} (\bibinfo {year}
  {2015})}\BibitemShut {NoStop}%
\bibitem [{\citenamefont {{Beik Mohammadi}}\ and\ \citenamefont
  {Kent}(2021)}]{BeikMohammadi2021}%
  \BibitemOpen
  \bibfield  {author} {\bibinfo {author} {\bibfnamefont {J.}~\bibnamefont
  {{Beik Mohammadi}}}\ and\ \bibinfo {author} {\bibfnamefont {A.~D.}\
  \bibnamefont {Kent}},\ }\bibfield  {title} {\enquote {\bibinfo {title}
  {{Spin-torque switching mechanisms of perpendicular magnetic tunnel junction
  nanopillars}},}\ }\href {\doibase 10.1063/5.0046596} {\bibfield  {journal}
  {\bibinfo  {journal} {Applied Physics Letters}\ }\textbf {\bibinfo {volume}
  {118}},\ \bibinfo {pages} {132407} (\bibinfo {year} {2021})},\ \Eprint
  {http://arxiv.org/abs/2003.13875} {arXiv:2003.13875} \BibitemShut {NoStop}%
\bibitem [{\citenamefont {Bouquin}\ \emph {et~al.}(2021)\citenamefont
  {Bouquin}, \citenamefont {Kim}, \citenamefont {Bultynck}, \citenamefont
  {Rao}, \citenamefont {Couet}, \citenamefont {Kar},\ and\ \citenamefont
  {Devolder}}]{Bouquin2021}%
  \BibitemOpen
  \bibfield  {author} {\bibinfo {author} {\bibfnamefont {P.}~\bibnamefont
  {Bouquin}}, \bibinfo {author} {\bibfnamefont {J.-V.}\ \bibnamefont {Kim}},
  \bibinfo {author} {\bibfnamefont {O.}~\bibnamefont {Bultynck}}, \bibinfo
  {author} {\bibfnamefont {S.}~\bibnamefont {Rao}}, \bibinfo {author}
  {\bibfnamefont {S.}~\bibnamefont {Couet}}, \bibinfo {author} {\bibfnamefont
  {G.~S.}\ \bibnamefont {Kar}}, \ and\ \bibinfo {author} {\bibfnamefont
  {T.}~\bibnamefont {Devolder}},\ }\bibfield  {title} {\enquote {\bibinfo
  {title} {{Stochastic Processes in Magnetization Reversal Involving
  Domain-Wall Motion in Magnetic Memory Elements}},}\ }\href {\doibase
  10.1103/PhysRevApplied.15.024037} {\bibfield  {journal} {\bibinfo  {journal}
  {Physical Review Applied}\ }\textbf {\bibinfo {volume} {15}},\ \bibinfo
  {pages} {024037} (\bibinfo {year} {2021})}\BibitemShut {NoStop}%
\bibitem [{\citenamefont {Devolder}\ \emph {et~al.}(2008)\citenamefont
  {Devolder}, \citenamefont {Hayakawa}, \citenamefont {Ito}, \citenamefont
  {Takahashi}, \citenamefont {Ikeda}, \citenamefont {Crozat}, \citenamefont
  {Zerounian}, \citenamefont {Kim}, \citenamefont {Chappert},\ and\
  \citenamefont {Ohno}}]{Devolder2008}%
  \BibitemOpen
  \bibfield  {author} {\bibinfo {author} {\bibfnamefont {T.}~\bibnamefont
  {Devolder}}, \bibinfo {author} {\bibfnamefont {J.}~\bibnamefont {Hayakawa}},
  \bibinfo {author} {\bibfnamefont {K.}~\bibnamefont {Ito}}, \bibinfo {author}
  {\bibfnamefont {H.}~\bibnamefont {Takahashi}}, \bibinfo {author}
  {\bibfnamefont {S.}~\bibnamefont {Ikeda}}, \bibinfo {author} {\bibfnamefont
  {P.}~\bibnamefont {Crozat}}, \bibinfo {author} {\bibfnamefont
  {N.}~\bibnamefont {Zerounian}}, \bibinfo {author} {\bibfnamefont {J.-V.}\
  \bibnamefont {Kim}}, \bibinfo {author} {\bibfnamefont {C.}~\bibnamefont
  {Chappert}}, \ and\ \bibinfo {author} {\bibfnamefont {H.}~\bibnamefont
  {Ohno}},\ }\bibfield  {title} {\enquote {\bibinfo {title} {{Single-Shot
  Time-Resolved Measurements of Nanosecond-Scale Spin-Transfer Induced
  Switching: Stochastic Versus Deterministic Aspects}},}\ }\href {\doibase
  10.1103/PhysRevLett.100.057206} {\bibfield  {journal} {\bibinfo  {journal}
  {Physical Review Letters}\ }\textbf {\bibinfo {volume} {100}},\ \bibinfo
  {pages} {057206} (\bibinfo {year} {2008})}\BibitemShut {NoStop}%
\bibitem [{\citenamefont {Tomita}\ \emph {et~al.}(2008)\citenamefont {Tomita},
  \citenamefont {Konishi}, \citenamefont {Nozaki}, \citenamefont {Kubota},
  \citenamefont {Fukushima}, \citenamefont {Yakushiji}, \citenamefont {Yuasa},
  \citenamefont {Nakatani}, \citenamefont {Shinjo}, \citenamefont {Shiraishi},\
  and\ \citenamefont {Suzuki}}]{Tomita2008}%
  \BibitemOpen
  \bibfield  {author} {\bibinfo {author} {\bibfnamefont {H.}~\bibnamefont
  {Tomita}}, \bibinfo {author} {\bibfnamefont {K.}~\bibnamefont {Konishi}},
  \bibinfo {author} {\bibfnamefont {T.}~\bibnamefont {Nozaki}}, \bibinfo
  {author} {\bibfnamefont {H.}~\bibnamefont {Kubota}}, \bibinfo {author}
  {\bibfnamefont {A.}~\bibnamefont {Fukushima}}, \bibinfo {author}
  {\bibfnamefont {K.}~\bibnamefont {Yakushiji}}, \bibinfo {author}
  {\bibfnamefont {S.}~\bibnamefont {Yuasa}}, \bibinfo {author} {\bibfnamefont
  {Y.}~\bibnamefont {Nakatani}}, \bibinfo {author} {\bibfnamefont
  {T.}~\bibnamefont {Shinjo}}, \bibinfo {author} {\bibfnamefont
  {M.}~\bibnamefont {Shiraishi}}, \ and\ \bibinfo {author} {\bibfnamefont
  {Y.}~\bibnamefont {Suzuki}},\ }\bibfield  {title} {\enquote {\bibinfo {title}
  {{Single-shot measurements of spin-transfer switching in CoFeB/ MgO/CoFeB
  magnetic tunnel junctions}},}\ }\href {\doibase 10.1143/APEX.1.061303}
  {\bibfield  {journal} {\bibinfo  {journal} {Applied Physics Express}\
  }\textbf {\bibinfo {volume} {1}},\ \bibinfo {pages} {0613031--0613033}
  (\bibinfo {year} {2008})}\BibitemShut {NoStop}%
\bibitem [{\citenamefont {Grimaldi}\ \emph {et~al.}(2020)\citenamefont
  {Grimaldi}, \citenamefont {Krizakova}, \citenamefont {Sala}, \citenamefont
  {Yasin}, \citenamefont {Couet}, \citenamefont {{Sankar Kar}}, \citenamefont
  {Garello},\ and\ \citenamefont {Gambardella}}]{Grimaldi2020}%
  \BibitemOpen
  \bibfield  {author} {\bibinfo {author} {\bibfnamefont {E.}~\bibnamefont
  {Grimaldi}}, \bibinfo {author} {\bibfnamefont {V.}~\bibnamefont {Krizakova}},
  \bibinfo {author} {\bibfnamefont {G.}~\bibnamefont {Sala}}, \bibinfo {author}
  {\bibfnamefont {F.}~\bibnamefont {Yasin}}, \bibinfo {author} {\bibfnamefont
  {S.}~\bibnamefont {Couet}}, \bibinfo {author} {\bibfnamefont
  {G.}~\bibnamefont {{Sankar Kar}}}, \bibinfo {author} {\bibfnamefont
  {K.}~\bibnamefont {Garello}}, \ and\ \bibinfo {author} {\bibfnamefont
  {P.}~\bibnamefont {Gambardella}},\ }\bibfield  {title} {\enquote {\bibinfo
  {title} {{Single-shot dynamics of spin–orbit torque and spin transfer
  torque switching in three-terminal magnetic tunnel junctions}},}\ }\href
  {\doibase 10.1038/s41565-019-0607-7} {\bibfield  {journal} {\bibinfo
  {journal} {Nature Nanotechnology}\ }\textbf {\bibinfo {volume} {15}},\
  \bibinfo {pages} {111--117} (\bibinfo {year} {2020})}\BibitemShut {NoStop}%
\bibitem [{\citenamefont {Krizakova}\ \emph {et~al.}(2020)\citenamefont
  {Krizakova}, \citenamefont {Garello}, \citenamefont {Grimaldi}, \citenamefont
  {Kar},\ and\ \citenamefont {Gambardella}}]{Krizakova2020}%
  \BibitemOpen
  \bibfield  {author} {\bibinfo {author} {\bibfnamefont {V.}~\bibnamefont
  {Krizakova}}, \bibinfo {author} {\bibfnamefont {K.}~\bibnamefont {Garello}},
  \bibinfo {author} {\bibfnamefont {E.}~\bibnamefont {Grimaldi}}, \bibinfo
  {author} {\bibfnamefont {G.~S.}\ \bibnamefont {Kar}}, \ and\ \bibinfo
  {author} {\bibfnamefont {P.}~\bibnamefont {Gambardella}},\ }\bibfield
  {title} {\enquote {\bibinfo {title} {{Field-free switching of magnetic tunnel
  junctions driven by spin–orbit torques at sub-ns timescales}},}\ }\href
  {\doibase 10.1063/5.0011433} {\bibfield  {journal} {\bibinfo  {journal}
  {Applied Physics Letters}\ }\textbf {\bibinfo {volume} {116}},\ \bibinfo
  {pages} {232406} (\bibinfo {year} {2020})},\ \Eprint
  {http://arxiv.org/abs/2006.06390} {arXiv:2006.06390} \BibitemShut {NoStop}%
\bibitem [{\citenamefont {Sala}\ \emph {et~al.}(2021)\citenamefont {Sala},
  \citenamefont {Krizakova}, \citenamefont {Grimaldi}, \citenamefont {Lambert},
  \citenamefont {Devolder},\ and\ \citenamefont {Gambardella}}]{Sala2021}%
  \BibitemOpen
  \bibfield  {author} {\bibinfo {author} {\bibfnamefont {G.}~\bibnamefont
  {Sala}}, \bibinfo {author} {\bibfnamefont {V.}~\bibnamefont {Krizakova}},
  \bibinfo {author} {\bibfnamefont {E.}~\bibnamefont {Grimaldi}}, \bibinfo
  {author} {\bibfnamefont {C.-H.}\ \bibnamefont {Lambert}}, \bibinfo {author}
  {\bibfnamefont {T.}~\bibnamefont {Devolder}}, \ and\ \bibinfo {author}
  {\bibfnamefont {P.}~\bibnamefont {Gambardella}},\ }\bibfield  {title}
  {\enquote {\bibinfo {title} {{Real-time Hall-effect detection of
  current-induced magnetization dynamics in ferrimagnets}},}\ }\href {\doibase
  10.1038/s41467-021-20968-0} {\bibfield  {journal} {\bibinfo  {journal}
  {Nature Communications}\ }\textbf {\bibinfo {volume} {12}},\ \bibinfo {pages}
  {656} (\bibinfo {year} {2021})}\BibitemShut {NoStop}%
\bibitem [{\citenamefont {Bultynck}\ \emph {et~al.}(2018)\citenamefont
  {Bultynck}, \citenamefont {Manfrini}, \citenamefont {Vaysset}, \citenamefont
  {Swerts}, \citenamefont {Wilson}, \citenamefont {Sor{\'{e}}e}, \citenamefont
  {Heyns}, \citenamefont {Mocuta}, \citenamefont {Radu},\ and\ \citenamefont
  {Devolder}}]{Bultynck2018}%
  \BibitemOpen
  \bibfield  {author} {\bibinfo {author} {\bibfnamefont {O.}~\bibnamefont
  {Bultynck}}, \bibinfo {author} {\bibfnamefont {M.}~\bibnamefont {Manfrini}},
  \bibinfo {author} {\bibfnamefont {A.}~\bibnamefont {Vaysset}}, \bibinfo
  {author} {\bibfnamefont {J.}~\bibnamefont {Swerts}}, \bibinfo {author}
  {\bibfnamefont {C.~J.}\ \bibnamefont {Wilson}}, \bibinfo {author}
  {\bibfnamefont {B.}~\bibnamefont {Sor{\'{e}}e}}, \bibinfo {author}
  {\bibfnamefont {M.}~\bibnamefont {Heyns}}, \bibinfo {author} {\bibfnamefont
  {D.}~\bibnamefont {Mocuta}}, \bibinfo {author} {\bibfnamefont {I.~P.}\
  \bibnamefont {Radu}}, \ and\ \bibinfo {author} {\bibfnamefont
  {T.}~\bibnamefont {Devolder}},\ }\bibfield  {title} {\enquote {\bibinfo
  {title} {{Instant-On Spin Torque in Noncollinear Magnetic Tunnel
  Junctions}},}\ }\href {\doibase 10.1103/PhysRevApplied.10.054028} {\bibfield
  {journal} {\bibinfo  {journal} {Physical Review Applied}\ }\textbf {\bibinfo
  {volume} {10}},\ \bibinfo {pages} {1} (\bibinfo {year} {2018})}\BibitemShut
  {NoStop}%
\bibitem [{\citenamefont {Koch}, \citenamefont {Katine},\ and\ \citenamefont
  {Sun}(2004)}]{Koch2004}%
  \BibitemOpen
  \bibfield  {author} {\bibinfo {author} {\bibfnamefont {R.~H.}\ \bibnamefont
  {Koch}}, \bibinfo {author} {\bibfnamefont {J.~A.}\ \bibnamefont {Katine}}, \
  and\ \bibinfo {author} {\bibfnamefont {J.~Z.}\ \bibnamefont {Sun}},\
  }\bibfield  {title} {\enquote {\bibinfo {title} {{Time-Resolved Reversal of
  Spin-Transfer Switching in a Nanomagnet}},}\ }\href {\doibase
  10.1103/PhysRevLett.92.088302} {\bibfield  {journal} {\bibinfo  {journal}
  {Physical Review Letters}\ }\textbf {\bibinfo {volume} {92}},\ \bibinfo
  {pages} {088302} (\bibinfo {year} {2004})}\BibitemShut {NoStop}%
\bibitem [{\citenamefont {Li}\ and\ \citenamefont {Zhang}(2004)}]{Li2004}%
  \BibitemOpen
  \bibfield  {author} {\bibinfo {author} {\bibfnamefont {Z.}~\bibnamefont
  {Li}}\ and\ \bibinfo {author} {\bibfnamefont {S.}~\bibnamefont {Zhang}},\
  }\bibfield  {title} {\enquote {\bibinfo {title} {{Thermally assisted
  magnetization reversal in the presence of a spin-transfer torque}},}\ }\href
  {\doibase 10.1103/PhysRevB.69.134416} {\bibfield  {journal} {\bibinfo
  {journal} {Physical Review B - Condensed Matter and Materials Physics}\
  }\textbf {\bibinfo {volume} {69}},\ \bibinfo {pages} {1--6} (\bibinfo {year}
  {2004})},\ \Eprint {http://arxiv.org/abs/0302339} {arXiv:0302339 [cond-mat]}
  \BibitemShut {NoStop}%
\bibitem [{\citenamefont {Bedau}\ \emph {et~al.}(2010)\citenamefont {Bedau},
  \citenamefont {Liu}, \citenamefont {Sun}, \citenamefont {Katine},
  \citenamefont {Fullerton}, \citenamefont {Mangin},\ and\ \citenamefont
  {Kent}}]{Bedau2010}%
  \BibitemOpen
  \bibfield  {author} {\bibinfo {author} {\bibfnamefont {D.}~\bibnamefont
  {Bedau}}, \bibinfo {author} {\bibfnamefont {H.}~\bibnamefont {Liu}}, \bibinfo
  {author} {\bibfnamefont {J.~Z.}\ \bibnamefont {Sun}}, \bibinfo {author}
  {\bibfnamefont {J.~A.}\ \bibnamefont {Katine}}, \bibinfo {author}
  {\bibfnamefont {E.~E.}\ \bibnamefont {Fullerton}}, \bibinfo {author}
  {\bibfnamefont {S.}~\bibnamefont {Mangin}}, \ and\ \bibinfo {author}
  {\bibfnamefont {A.~D.}\ \bibnamefont {Kent}},\ }\bibfield  {title} {\enquote
  {\bibinfo {title} {{Spin-transfer pulse switching: From the dynamic to the
  thermally activated regime}},}\ }\href {\doibase 10.1063/1.3532960}
  {\bibfield  {journal} {\bibinfo  {journal} {Applied Physics Letters}\
  }\textbf {\bibinfo {volume} {97}},\ \bibinfo {pages} {262502} (\bibinfo
  {year} {2010})}\BibitemShut {NoStop}%
\bibitem [{\citenamefont {Desplat}\ and\ \citenamefont
  {Kim}(2020)}]{Desplat2020}%
  \BibitemOpen
  \bibfield  {author} {\bibinfo {author} {\bibfnamefont {L.}~\bibnamefont
  {Desplat}}\ and\ \bibinfo {author} {\bibfnamefont {J.-V.}\ \bibnamefont
  {Kim}},\ }\bibfield  {title} {\enquote {\bibinfo {title} {{Entropy-reduced
  Retention Times in Magnetic Memory Elements: A Case of the Meyer-Neldel
  Compensation Rule}},}\ }\href {\doibase 10.1103/PhysRevLett.125.107201}
  {\bibfield  {journal} {\bibinfo  {journal} {Physical Review Letters}\
  }\textbf {\bibinfo {volume} {125}},\ \bibinfo {pages} {107201} (\bibinfo
  {year} {2020})},\ \Eprint {http://arxiv.org/abs/2007.02152}
  {arXiv:2007.02152} \BibitemShut {NoStop}%
\bibitem [{\citenamefont {Iga}\ \emph {et~al.}(2012)\citenamefont {Iga},
  \citenamefont {Yoshida}, \citenamefont {Ikeda}, \citenamefont {Hanyu},
  \citenamefont {Ohno},\ and\ \citenamefont {Endoh}}]{Iga2012}%
  \BibitemOpen
  \bibfield  {author} {\bibinfo {author} {\bibfnamefont {F.}~\bibnamefont
  {Iga}}, \bibinfo {author} {\bibfnamefont {Y.}~\bibnamefont {Yoshida}},
  \bibinfo {author} {\bibfnamefont {S.}~\bibnamefont {Ikeda}}, \bibinfo
  {author} {\bibfnamefont {T.}~\bibnamefont {Hanyu}}, \bibinfo {author}
  {\bibfnamefont {H.}~\bibnamefont {Ohno}}, \ and\ \bibinfo {author}
  {\bibfnamefont {T.}~\bibnamefont {Endoh}},\ }\bibfield  {title} {\enquote
  {\bibinfo {title} {{Time-Resolved Switching Characteristic in Magnetic Tunnel
  Junction with Spin Transfer Torque Write Scheme}},}\ }\href {\doibase
  10.1143/JJAP.51.02BM02} {\bibfield  {journal} {\bibinfo  {journal} {Japanese
  Journal of Applied Physics}\ }\textbf {\bibinfo {volume} {51}},\ \bibinfo
  {pages} {02BM02} (\bibinfo {year} {2012})}\BibitemShut {NoStop}%
\bibitem [{\citenamefont {Hahn}\ \emph {et~al.}(2016)\citenamefont {Hahn},
  \citenamefont {Wolf}, \citenamefont {Kardasz}, \citenamefont {Watts},
  \citenamefont {Pinarbasi},\ and\ \citenamefont {Kent}}]{Hahn2016}%
  \BibitemOpen
  \bibfield  {author} {\bibinfo {author} {\bibfnamefont {C.}~\bibnamefont
  {Hahn}}, \bibinfo {author} {\bibfnamefont {G.}~\bibnamefont {Wolf}}, \bibinfo
  {author} {\bibfnamefont {B.}~\bibnamefont {Kardasz}}, \bibinfo {author}
  {\bibfnamefont {S.}~\bibnamefont {Watts}}, \bibinfo {author} {\bibfnamefont
  {M.}~\bibnamefont {Pinarbasi}}, \ and\ \bibinfo {author} {\bibfnamefont
  {A.~D.}\ \bibnamefont {Kent}},\ }\bibfield  {title} {\enquote {\bibinfo
  {title} {{Time-resolved studies of the spin-transfer reversal mechanism in
  perpendicularly magnetized magnetic tunnel junctions}},}\ }\href {\doibase
  10.1103/PhysRevB.94.214432} {\bibfield  {journal} {\bibinfo  {journal}
  {Physical Review B}\ }\textbf {\bibinfo {volume} {94}},\ \bibinfo {pages}
  {214432} (\bibinfo {year} {2016})},\ \Eprint
  {http://arxiv.org/abs/1610.09710} {arXiv:1610.09710} \BibitemShut {NoStop}%
\bibitem [{\citenamefont {Visscher}, \citenamefont {Munira},\ and\
  \citenamefont {Rosati}(2016)}]{Visscher2016}%
  \BibitemOpen
  \bibfield  {author} {\bibinfo {author} {\bibfnamefont {P.~B.}\ \bibnamefont
  {Visscher}}, \bibinfo {author} {\bibfnamefont {K.}~\bibnamefont {Munira}}, \
  and\ \bibinfo {author} {\bibfnamefont {R.~J.}\ \bibnamefont {Rosati}},\
  }\bibfield  {title} {\enquote {\bibinfo {title} {{Instability Mechanism for
  STT-MRAM switching}},}\ }\href {http://arxiv.org/abs/1604.03992} {\ ,\
  \bibinfo {pages} {3--6} (\bibinfo {year} {2016})},\ \Eprint
  {http://arxiv.org/abs/1604.03992} {arXiv:1604.03992} \BibitemShut {NoStop}%
\bibitem [{\citenamefont {Devolder}\ \emph {et~al.}(2018)\citenamefont
  {Devolder}, \citenamefont {Kim}, \citenamefont {Swerts}, \citenamefont
  {Couet}, \citenamefont {Rao}, \citenamefont {Kim}, \citenamefont {Mertens},
  \citenamefont {Kar},\ and\ \citenamefont {Nikitin}}]{Devolder2018}%
  \BibitemOpen
  \bibfield  {author} {\bibinfo {author} {\bibfnamefont {T.}~\bibnamefont
  {Devolder}}, \bibinfo {author} {\bibfnamefont {J.-V.}\ \bibnamefont {Kim}},
  \bibinfo {author} {\bibfnamefont {J.}~\bibnamefont {Swerts}}, \bibinfo
  {author} {\bibfnamefont {S.}~\bibnamefont {Couet}}, \bibinfo {author}
  {\bibfnamefont {S.}~\bibnamefont {Rao}}, \bibinfo {author} {\bibfnamefont
  {W.}~\bibnamefont {Kim}}, \bibinfo {author} {\bibfnamefont {S.}~\bibnamefont
  {Mertens}}, \bibinfo {author} {\bibfnamefont {G.}~\bibnamefont {Kar}}, \ and\
  \bibinfo {author} {\bibfnamefont {V.}~\bibnamefont {Nikitin}},\ }\bibfield
  {title} {\enquote {\bibinfo {title} {{Material Developments and Domain
  Wall-Based Nanosecond-Scale Switching Process in Perpendicularly Magnetized
  STT-MRAM Cells}},}\ }\href {\doibase 10.1109/TMAG.2017.2739187} {\bibfield
  {journal} {\bibinfo  {journal} {IEEE Transactions on Magnetics}\ }\textbf
  {\bibinfo {volume} {54}},\ \bibinfo {pages} {1--9} (\bibinfo {year}
  {2018})},\ \Eprint {http://arxiv.org/abs/1703.03198} {arXiv:1703.03198}
  \BibitemShut {NoStop}%
\bibitem [{\citenamefont {Munira}\ and\ \citenamefont
  {Visscher}(2015)}]{Munira2015}%
  \BibitemOpen
  \bibfield  {author} {\bibinfo {author} {\bibfnamefont {K.}~\bibnamefont
  {Munira}}\ and\ \bibinfo {author} {\bibfnamefont {P.~B.}\ \bibnamefont
  {Visscher}},\ }\bibfield  {title} {\enquote {\bibinfo {title} {{Calculation
  of energy-barrier lowering by incoherent switching in spin-transfer torque
  magnetoresistive random-access memory}},}\ }\href {\doibase
  10.1063/1.4908153} {\bibfield  {journal} {\bibinfo  {journal} {Journal of
  Applied Physics}\ }\textbf {\bibinfo {volume} {117}} (\bibinfo {year}
  {2015}),\ 10.1063/1.4908153}\BibitemShut {NoStop}%
\bibitem [{\citenamefont {Liu}\ \emph {et~al.}(2014)\citenamefont {Liu},
  \citenamefont {Bedau}, \citenamefont {Sun}, \citenamefont {Mangin},
  \citenamefont {Fullerton}, \citenamefont {Katine},\ and\ \citenamefont
  {Kent}}]{Liu2014b}%
  \BibitemOpen
  \bibfield  {author} {\bibinfo {author} {\bibfnamefont {H.}~\bibnamefont
  {Liu}}, \bibinfo {author} {\bibfnamefont {D.}~\bibnamefont {Bedau}}, \bibinfo
  {author} {\bibfnamefont {J.~Z.}\ \bibnamefont {Sun}}, \bibinfo {author}
  {\bibfnamefont {S.}~\bibnamefont {Mangin}}, \bibinfo {author} {\bibfnamefont
  {E.~E.}\ \bibnamefont {Fullerton}}, \bibinfo {author} {\bibfnamefont {J.~A.}\
  \bibnamefont {Katine}}, \ and\ \bibinfo {author} {\bibfnamefont {A.~D.}\
  \bibnamefont {Kent}},\ }\bibfield  {title} {\enquote {\bibinfo {title}
  {{Dynamics of spin torque switching in all-perpendicular spin valve
  nanopillars}},}\ }\href {\doibase 10.1016/j.jmmm.2014.01.061} {\bibfield
  {journal} {\bibinfo  {journal} {Journal of Magnetism and Magnetic Materials}\
  }\textbf {\bibinfo {volume} {358-359}},\ \bibinfo {pages} {233--258}
  (\bibinfo {year} {2014})}\BibitemShut {NoStop}%
\bibitem [{\citenamefont {Garello}\ \emph {et~al.}(2014)\citenamefont
  {Garello}, \citenamefont {Avci}, \citenamefont {Miron}, \citenamefont
  {Baumgartner}, \citenamefont {Ghosh}, \citenamefont {Auffret}, \citenamefont
  {Boulle}, \citenamefont {Gaudin},\ and\ \citenamefont
  {Gambardella}}]{Garello2014}%
  \BibitemOpen
  \bibfield  {author} {\bibinfo {author} {\bibfnamefont {K.}~\bibnamefont
  {Garello}}, \bibinfo {author} {\bibfnamefont {C.~O.}\ \bibnamefont {Avci}},
  \bibinfo {author} {\bibfnamefont {I.~M.}\ \bibnamefont {Miron}}, \bibinfo
  {author} {\bibfnamefont {M.}~\bibnamefont {Baumgartner}}, \bibinfo {author}
  {\bibfnamefont {A.}~\bibnamefont {Ghosh}}, \bibinfo {author} {\bibfnamefont
  {S.}~\bibnamefont {Auffret}}, \bibinfo {author} {\bibfnamefont
  {O.}~\bibnamefont {Boulle}}, \bibinfo {author} {\bibfnamefont
  {G.}~\bibnamefont {Gaudin}}, \ and\ \bibinfo {author} {\bibfnamefont
  {P.}~\bibnamefont {Gambardella}},\ }\bibfield  {title} {\enquote {\bibinfo
  {title} {{Ultrafast magnetization switching by spin-orbit torques}},}\ }\href
  {\doibase 10.1063/1.4902443} {\bibfield  {journal} {\bibinfo  {journal}
  {Applied Physics Letters}\ }\textbf {\bibinfo {volume} {105}},\ \bibinfo
  {pages} {212402} (\bibinfo {year} {2014})},\ \Eprint
  {http://arxiv.org/abs/1310.5586} {arXiv:1310.5586} \BibitemShut {NoStop}%
\bibitem [{\citenamefont {Cai}\ \emph {et~al.}(2021)\citenamefont {Cai},
  \citenamefont {Shi}, \citenamefont {Zhuo}, \citenamefont {Zhu}, \citenamefont
  {Huang}, \citenamefont {Yin}, \citenamefont {Cao}, \citenamefont {Wang},
  \citenamefont {Guo}, \citenamefont {Wang}, \citenamefont {Wang},\ and\
  \citenamefont {Zhao}}]{Cai2021}%
  \BibitemOpen
  \bibfield  {author} {\bibinfo {author} {\bibfnamefont {W.}~\bibnamefont
  {Cai}}, \bibinfo {author} {\bibfnamefont {K.}~\bibnamefont {Shi}}, \bibinfo
  {author} {\bibfnamefont {Y.}~\bibnamefont {Zhuo}}, \bibinfo {author}
  {\bibfnamefont {D.}~\bibnamefont {Zhu}}, \bibinfo {author} {\bibfnamefont
  {Y.}~\bibnamefont {Huang}}, \bibinfo {author} {\bibfnamefont
  {J.}~\bibnamefont {Yin}}, \bibinfo {author} {\bibfnamefont {K.}~\bibnamefont
  {Cao}}, \bibinfo {author} {\bibfnamefont {Z.}~\bibnamefont {Wang}}, \bibinfo
  {author} {\bibfnamefont {Z.}~\bibnamefont {Guo}}, \bibinfo {author}
  {\bibfnamefont {Z.}~\bibnamefont {Wang}}, \bibinfo {author} {\bibfnamefont
  {G.}~\bibnamefont {Wang}}, \ and\ \bibinfo {author} {\bibfnamefont
  {W.}~\bibnamefont {Zhao}},\ }\bibfield  {title} {\enquote {\bibinfo {title}
  {{Sub-ns Field-Free Switching in Perpendicular Magnetic Tunnel Junctions by
  the Interplay of Spin Transfer and Orbit Torques}},}\ }\href {\doibase
  10.1109/LED.2021.3069391} {\bibfield  {journal} {\bibinfo  {journal} {IEEE
  Electron Device Letters}\ }\textbf {\bibinfo {volume} {42}},\ \bibinfo
  {pages} {704--707} (\bibinfo {year} {2021})}\BibitemShut {NoStop}%
\bibitem [{\citenamefont {Mikuszeit}\ \emph {et~al.}(2015)\citenamefont
  {Mikuszeit}, \citenamefont {Boulle}, \citenamefont {Miron}, \citenamefont
  {Garello}, \citenamefont {Gambardella}, \citenamefont {Gaudin},\ and\
  \citenamefont {Buda-Prejbeanu}}]{Mikuszeit2015}%
  \BibitemOpen
  \bibfield  {author} {\bibinfo {author} {\bibfnamefont {N.}~\bibnamefont
  {Mikuszeit}}, \bibinfo {author} {\bibfnamefont {O.}~\bibnamefont {Boulle}},
  \bibinfo {author} {\bibfnamefont {I.~M.}\ \bibnamefont {Miron}}, \bibinfo
  {author} {\bibfnamefont {K.}~\bibnamefont {Garello}}, \bibinfo {author}
  {\bibfnamefont {P.}~\bibnamefont {Gambardella}}, \bibinfo {author}
  {\bibfnamefont {G.}~\bibnamefont {Gaudin}}, \ and\ \bibinfo {author}
  {\bibfnamefont {L.~D.}\ \bibnamefont {Buda-Prejbeanu}},\ }\bibfield  {title}
  {\enquote {\bibinfo {title} {{Spin-orbit torque driven chiral magnetization
  reversal in ultrathin nanostructures}},}\ }\href {\doibase
  10.1103/PhysRevB.92.144424} {\bibfield  {journal} {\bibinfo  {journal}
  {Physical Review B}\ }\textbf {\bibinfo {volume} {92}},\ \bibinfo {pages}
  {144424} (\bibinfo {year} {2015})},\ \Eprint
  {http://arxiv.org/abs/1509.07341} {arXiv:1509.07341} \BibitemShut {NoStop}%
\bibitem [{\citenamefont {Martinez}\ \emph {et~al.}(2015)\citenamefont
  {Martinez}, \citenamefont {Torres}, \citenamefont {Perez}, \citenamefont
  {Hernandez}, \citenamefont {Raposo},\ and\ \citenamefont
  {Moretti}}]{Martinez2015}%
  \BibitemOpen
  \bibfield  {author} {\bibinfo {author} {\bibfnamefont {E.}~\bibnamefont
  {Martinez}}, \bibinfo {author} {\bibfnamefont {L.}~\bibnamefont {Torres}},
  \bibinfo {author} {\bibfnamefont {N.}~\bibnamefont {Perez}}, \bibinfo
  {author} {\bibfnamefont {M.~A.}\ \bibnamefont {Hernandez}}, \bibinfo {author}
  {\bibfnamefont {V.}~\bibnamefont {Raposo}}, \ and\ \bibinfo {author}
  {\bibfnamefont {S.}~\bibnamefont {Moretti}},\ }\bibfield  {title} {\enquote
  {\bibinfo {title} {{Universal chiral-triggered magnetization switching in
  confined nanodots}},}\ }\href {\doibase 10.1038/srep10156} {\bibfield
  {journal} {\bibinfo  {journal} {Scientific Reports}\ }\textbf {\bibinfo
  {volume} {5}},\ \bibinfo {pages} {1--15} (\bibinfo {year}
  {2015})}\BibitemShut {NoStop}%
\bibitem [{\citenamefont {Sala}\ \emph {et~al.}(2022)\citenamefont {Sala},
  \citenamefont {Lambert}, \citenamefont {Finizio}, \citenamefont {Raposo},
  \citenamefont {Krizakova}, \citenamefont {Krishnaswamy}, \citenamefont
  {Weigand}, \citenamefont {Raabe}, \citenamefont {Rossell}, \citenamefont
  {Martinez},\ and\ \citenamefont {Gambardella}}]{Sala2022}%
  \BibitemOpen
  \bibfield  {author} {\bibinfo {author} {\bibfnamefont {G.}~\bibnamefont
  {Sala}}, \bibinfo {author} {\bibfnamefont {C.-h.}\ \bibnamefont {Lambert}},
  \bibinfo {author} {\bibfnamefont {S.}~\bibnamefont {Finizio}}, \bibinfo
  {author} {\bibfnamefont {V.}~\bibnamefont {Raposo}}, \bibinfo {author}
  {\bibfnamefont {V.}~\bibnamefont {Krizakova}}, \bibinfo {author}
  {\bibfnamefont {G.}~\bibnamefont {Krishnaswamy}}, \bibinfo {author}
  {\bibfnamefont {M.}~\bibnamefont {Weigand}}, \bibinfo {author} {\bibfnamefont
  {J.}~\bibnamefont {Raabe}}, \bibinfo {author} {\bibfnamefont {M.~D.}\
  \bibnamefont {Rossell}}, \bibinfo {author} {\bibfnamefont {E.}~\bibnamefont
  {Martinez}}, \ and\ \bibinfo {author} {\bibfnamefont {P.}~\bibnamefont
  {Gambardella}},\ }\bibfield  {title} {\enquote {\bibinfo {title}
  {{Asynchronous current-induced switching of rare-earth and transition-metal
  sublattices in ferrimagnetic alloys}},}\ }\href {\doibase
  10.1038/s41563-022-01248-8} {\bibfield  {journal} {\bibinfo  {journal}
  {Nature Materials}\ }\textbf {\bibinfo {volume} {21}},\ \bibinfo {pages}
  {640--646} (\bibinfo {year} {2022})}\BibitemShut {NoStop}%
\bibitem [{\citenamefont {Pizzini}\ \emph {et~al.}(2014)\citenamefont
  {Pizzini}, \citenamefont {Vogel}, \citenamefont {Rohart}, \citenamefont
  {Buda-Prejbeanu}, \citenamefont {Ju{\'{e}}}, \citenamefont {Boulle},
  \citenamefont {Miron}, \citenamefont {Safeer}, \citenamefont {Auffret},
  \citenamefont {Gaudin},\ and\ \citenamefont {Thiaville}}]{Pizzini2014}%
  \BibitemOpen
  \bibfield  {author} {\bibinfo {author} {\bibfnamefont {S.}~\bibnamefont
  {Pizzini}}, \bibinfo {author} {\bibfnamefont {J.}~\bibnamefont {Vogel}},
  \bibinfo {author} {\bibfnamefont {S.}~\bibnamefont {Rohart}}, \bibinfo
  {author} {\bibfnamefont {L.~D.}\ \bibnamefont {Buda-Prejbeanu}}, \bibinfo
  {author} {\bibfnamefont {E.}~\bibnamefont {Ju{\'{e}}}}, \bibinfo {author}
  {\bibfnamefont {O.}~\bibnamefont {Boulle}}, \bibinfo {author} {\bibfnamefont
  {I.~M.}\ \bibnamefont {Miron}}, \bibinfo {author} {\bibfnamefont {C.~K.}\
  \bibnamefont {Safeer}}, \bibinfo {author} {\bibfnamefont {S.}~\bibnamefont
  {Auffret}}, \bibinfo {author} {\bibfnamefont {G.}~\bibnamefont {Gaudin}}, \
  and\ \bibinfo {author} {\bibfnamefont {A.}~\bibnamefont {Thiaville}},\
  }\bibfield  {title} {\enquote {\bibinfo {title} {{Chirality-Induced
  Asymmetric Magnetic Nucleation in Pt/Co/AlOx Ultrathin Microstructures}},}\
  }\href {\doibase 10.1103/PhysRevLett.113.047203} {\bibfield  {journal}
  {\bibinfo  {journal} {Physical Review Letters}\ }\textbf {\bibinfo {volume}
  {113}},\ \bibinfo {pages} {047203} (\bibinfo {year} {2014})},\ \Eprint
  {http://arxiv.org/abs/1403.4694} {arXiv:1403.4694} \BibitemShut {NoStop}%
\bibitem [{\citenamefont {Jhuria}\ \emph {et~al.}(2020)\citenamefont {Jhuria},
  \citenamefont {Hohlfeld}, \citenamefont {Pattabi}, \citenamefont {Martin},
  \citenamefont {{Arriola C{\'{o}}rdova}}, \citenamefont {Shi}, \citenamefont
  {{Lo Conte}}, \citenamefont {Petit-Watelot}, \citenamefont {Rojas-Sanchez},
  \citenamefont {Malinowski}, \citenamefont {Mangin}, \citenamefont
  {Lema{\^{i}}tre}, \citenamefont {Hehn}, \citenamefont {Bokor}, \citenamefont
  {Wilson},\ and\ \citenamefont {Gorchon}}]{Jhuria2020}%
  \BibitemOpen
  \bibfield  {author} {\bibinfo {author} {\bibfnamefont {K.}~\bibnamefont
  {Jhuria}}, \bibinfo {author} {\bibfnamefont {J.}~\bibnamefont {Hohlfeld}},
  \bibinfo {author} {\bibfnamefont {A.}~\bibnamefont {Pattabi}}, \bibinfo
  {author} {\bibfnamefont {E.}~\bibnamefont {Martin}}, \bibinfo {author}
  {\bibfnamefont {A.~Y.}\ \bibnamefont {{Arriola C{\'{o}}rdova}}}, \bibinfo
  {author} {\bibfnamefont {X.}~\bibnamefont {Shi}}, \bibinfo {author}
  {\bibfnamefont {R.}~\bibnamefont {{Lo Conte}}}, \bibinfo {author}
  {\bibfnamefont {S.}~\bibnamefont {Petit-Watelot}}, \bibinfo {author}
  {\bibfnamefont {J.~C.}\ \bibnamefont {Rojas-Sanchez}}, \bibinfo {author}
  {\bibfnamefont {G.}~\bibnamefont {Malinowski}}, \bibinfo {author}
  {\bibfnamefont {S.}~\bibnamefont {Mangin}}, \bibinfo {author} {\bibfnamefont
  {A.}~\bibnamefont {Lema{\^{i}}tre}}, \bibinfo {author} {\bibfnamefont
  {M.}~\bibnamefont {Hehn}}, \bibinfo {author} {\bibfnamefont {J.}~\bibnamefont
  {Bokor}}, \bibinfo {author} {\bibfnamefont {R.~B.}\ \bibnamefont {Wilson}}, \
  and\ \bibinfo {author} {\bibfnamefont {J.}~\bibnamefont {Gorchon}},\
  }\bibfield  {title} {\enquote {\bibinfo {title} {{Spin-orbit torque switching
  of a ferromagnet with picosecond electrical pulses}},}\ }\href {\doibase
  10.1038/s41928-020-00488-3} {\bibfield  {journal} {\bibinfo  {journal}
  {Nature Electronics}\ }\textbf {\bibinfo {volume} {3}},\ \bibinfo {pages}
  {680--686} (\bibinfo {year} {2020})}\BibitemShut {NoStop}%
\bibitem [{\citenamefont {Yang}\ \emph {et~al.}(2017)\citenamefont {Yang},
  \citenamefont {Wilson}, \citenamefont {Gorchon}, \citenamefont {Lambert},
  \citenamefont {Salahuddin},\ and\ \citenamefont {Bokor}}]{Yang2017}%
  \BibitemOpen
  \bibfield  {author} {\bibinfo {author} {\bibfnamefont {Y.}~\bibnamefont
  {Yang}}, \bibinfo {author} {\bibfnamefont {R.~B.}\ \bibnamefont {Wilson}},
  \bibinfo {author} {\bibfnamefont {J.}~\bibnamefont {Gorchon}}, \bibinfo
  {author} {\bibfnamefont {C.~H.}\ \bibnamefont {Lambert}}, \bibinfo {author}
  {\bibfnamefont {S.}~\bibnamefont {Salahuddin}}, \ and\ \bibinfo {author}
  {\bibfnamefont {J.}~\bibnamefont {Bokor}},\ }\bibfield  {title} {\enquote
  {\bibinfo {title} {{Ultrafast magnetization reversal by picosecond electrical
  pulses}},}\ }\href {\doibase 10.1126/sciadv.1603117} {\bibfield  {journal}
  {\bibinfo  {journal} {Science Advances}\ }\textbf {\bibinfo {volume} {3}},\
  \bibinfo {pages} {1--7} (\bibinfo {year} {2017})}\BibitemShut {NoStop}%
\bibitem [{\citenamefont {Wilson}\ \emph {et~al.}(2017)\citenamefont {Wilson},
  \citenamefont {Gorchon}, \citenamefont {Yang}, \citenamefont {Lambert},
  \citenamefont {Salahuddin},\ and\ \citenamefont {Bokor}}]{Wilson2017}%
  \BibitemOpen
  \bibfield  {author} {\bibinfo {author} {\bibfnamefont {R.~B.}\ \bibnamefont
  {Wilson}}, \bibinfo {author} {\bibfnamefont {J.}~\bibnamefont {Gorchon}},
  \bibinfo {author} {\bibfnamefont {Y.}~\bibnamefont {Yang}}, \bibinfo {author}
  {\bibfnamefont {C.~H.}\ \bibnamefont {Lambert}}, \bibinfo {author}
  {\bibfnamefont {S.}~\bibnamefont {Salahuddin}}, \ and\ \bibinfo {author}
  {\bibfnamefont {J.}~\bibnamefont {Bokor}},\ }\bibfield  {title} {\enquote
  {\bibinfo {title} {{Ultrafast magnetic switching of GdFeCo with electronic
  heat currents}},}\ }\href {\doibase 10.1103/PhysRevB.95.180409} {\bibfield
  {journal} {\bibinfo  {journal} {Physical Review B}\ }\textbf {\bibinfo
  {volume} {95}},\ \bibinfo {pages} {1--5} (\bibinfo {year}
  {2017})}\BibitemShut {NoStop}%
\bibitem [{\citenamefont {Cui}\ \emph {et~al.}(2010)\citenamefont {Cui},
  \citenamefont {Finocchio}, \citenamefont {Wang}, \citenamefont {Katine},
  \citenamefont {Buhrman},\ and\ \citenamefont {Ralph}}]{Cui2010}%
  \BibitemOpen
  \bibfield  {author} {\bibinfo {author} {\bibfnamefont {Y.~T.}\ \bibnamefont
  {Cui}}, \bibinfo {author} {\bibfnamefont {G.}~\bibnamefont {Finocchio}},
  \bibinfo {author} {\bibfnamefont {C.}~\bibnamefont {Wang}}, \bibinfo {author}
  {\bibfnamefont {J.~A.}\ \bibnamefont {Katine}}, \bibinfo {author}
  {\bibfnamefont {R.~A.}\ \bibnamefont {Buhrman}}, \ and\ \bibinfo {author}
  {\bibfnamefont {D.~C.}\ \bibnamefont {Ralph}},\ }\bibfield  {title} {\enquote
  {\bibinfo {title} {{Single-shot time-domain studies of spin-torque-driven
  switching in magnetic tunnel junctions}},}\ }\href {\doibase
  10.1103/PhysRevLett.104.097201} {\bibfield  {journal} {\bibinfo  {journal}
  {Physical Review Letters}\ }\textbf {\bibinfo {volume} {104}},\ \bibinfo
  {pages} {1--4} (\bibinfo {year} {2010})}\BibitemShut {NoStop}%
\bibitem [{\citenamefont {Zhao}\ \emph {et~al.}(2012)\citenamefont {Zhao},
  \citenamefont {Zhang}, \citenamefont {Amiri}, \citenamefont {Katine},
  \citenamefont {Langer}, \citenamefont {Jiang}, \citenamefont {Krivorotov},
  \citenamefont {Wang},\ and\ \citenamefont {Wang}}]{Zhao2012}%
  \BibitemOpen
  \bibfield  {author} {\bibinfo {author} {\bibfnamefont {H.}~\bibnamefont
  {Zhao}}, \bibinfo {author} {\bibfnamefont {Y.}~\bibnamefont {Zhang}},
  \bibinfo {author} {\bibfnamefont {P.~K.}\ \bibnamefont {Amiri}}, \bibinfo
  {author} {\bibfnamefont {J.~A.}\ \bibnamefont {Katine}}, \bibinfo {author}
  {\bibfnamefont {J.}~\bibnamefont {Langer}}, \bibinfo {author} {\bibfnamefont
  {H.}~\bibnamefont {Jiang}}, \bibinfo {author} {\bibfnamefont {I.~N.}\
  \bibnamefont {Krivorotov}}, \bibinfo {author} {\bibfnamefont {K.~L.}\
  \bibnamefont {Wang}}, \ and\ \bibinfo {author} {\bibfnamefont {J.-P.}\
  \bibnamefont {Wang}},\ }\bibfield  {title} {\enquote {\bibinfo {title}
  {{Spin-Torque Driven Switching Probability Density Function Asymmetry}},}\
  }\href {\doibase 10.1109/TMAG.2012.2197815} {\bibfield  {journal} {\bibinfo
  {journal} {IEEE Transactions on Magnetics}\ }\textbf {\bibinfo {volume}
  {48}},\ \bibinfo {pages} {3818--3820} (\bibinfo {year} {2012})}\BibitemShut
  {NoStop}%
\bibitem [{\citenamefont {Krizakova}\ \emph {et~al.}(2021)\citenamefont
  {Krizakova}, \citenamefont {Grimaldi}, \citenamefont {Garello}, \citenamefont
  {Sala}, \citenamefont {Couet}, \citenamefont {Kar},\ and\ \citenamefont
  {Gambardella}}]{Krizakova2021}%
  \BibitemOpen
  \bibfield  {author} {\bibinfo {author} {\bibfnamefont {V.}~\bibnamefont
  {Krizakova}}, \bibinfo {author} {\bibfnamefont {E.}~\bibnamefont {Grimaldi}},
  \bibinfo {author} {\bibfnamefont {K.}~\bibnamefont {Garello}}, \bibinfo
  {author} {\bibfnamefont {G.}~\bibnamefont {Sala}}, \bibinfo {author}
  {\bibfnamefont {S.}~\bibnamefont {Couet}}, \bibinfo {author} {\bibfnamefont
  {G.~S.}\ \bibnamefont {Kar}}, \ and\ \bibinfo {author} {\bibfnamefont
  {P.}~\bibnamefont {Gambardella}},\ }\bibfield  {title} {\enquote {\bibinfo
  {title} {{Interplay of Voltage Control of Magnetic Anisotropy, Spin-Transfer
  Torque, and Heat in the Spin-Orbit-Torque Switching of Three-Terminal
  Magnetic Tunnel Junctions}},}\ }\href {\doibase
  10.1103/PhysRevApplied.15.054055} {\bibfield  {journal} {\bibinfo  {journal}
  {Physical Review Applied}\ }\textbf {\bibinfo {volume} {15}},\ \bibinfo
  {pages} {1} (\bibinfo {year} {2021})},\ \Eprint
  {http://arxiv.org/abs/2106.01054} {arXiv:2106.01054} \BibitemShut {NoStop}%
\bibitem [{\citenamefont {Diao}\ \emph {et~al.}(2007)\citenamefont {Diao},
  \citenamefont {Li}, \citenamefont {Wang}, \citenamefont {Ding}, \citenamefont
  {Panchula}, \citenamefont {Chen}, \citenamefont {Wang},\ and\ \citenamefont
  {Huai}}]{Diao2007}%
  \BibitemOpen
  \bibfield  {author} {\bibinfo {author} {\bibfnamefont {Z.}~\bibnamefont
  {Diao}}, \bibinfo {author} {\bibfnamefont {Z.}~\bibnamefont {Li}}, \bibinfo
  {author} {\bibfnamefont {S.}~\bibnamefont {Wang}}, \bibinfo {author}
  {\bibfnamefont {Y.}~\bibnamefont {Ding}}, \bibinfo {author} {\bibfnamefont
  {A.}~\bibnamefont {Panchula}}, \bibinfo {author} {\bibfnamefont
  {E.}~\bibnamefont {Chen}}, \bibinfo {author} {\bibfnamefont {L.~C.}\
  \bibnamefont {Wang}}, \ and\ \bibinfo {author} {\bibfnamefont
  {Y.}~\bibnamefont {Huai}},\ }\bibfield  {title} {\enquote {\bibinfo {title}
  {{Spin-transfer torque switching in magnetic tunnel junctions and
  spin-transfer torque random access memory}},}\ }\href {\doibase
  10.1088/0953-8984/19/16/165209} {\bibfield  {journal} {\bibinfo  {journal}
  {Journal of Physics Condensed Matter}\ }\textbf {\bibinfo {volume} {19}}
  (\bibinfo {year} {2007}),\ 10.1088/0953-8984/19/16/165209}\BibitemShut
  {NoStop}%
\bibitem [{\citenamefont {Vincent}\ \emph {et~al.}(2015)\citenamefont
  {Vincent}, \citenamefont {Locatelli}, \citenamefont {Klein}, \citenamefont
  {Zhao}, \citenamefont {Galdin-Retailleau},\ and\ \citenamefont
  {Querlioz}}]{Vincent2015}%
  \BibitemOpen
  \bibfield  {author} {\bibinfo {author} {\bibfnamefont {A.~F.}\ \bibnamefont
  {Vincent}}, \bibinfo {author} {\bibfnamefont {N.}~\bibnamefont {Locatelli}},
  \bibinfo {author} {\bibfnamefont {J.~O.}\ \bibnamefont {Klein}}, \bibinfo
  {author} {\bibfnamefont {W.~S.}\ \bibnamefont {Zhao}}, \bibinfo {author}
  {\bibfnamefont {S.}~\bibnamefont {Galdin-Retailleau}}, \ and\ \bibinfo
  {author} {\bibfnamefont {D.}~\bibnamefont {Querlioz}},\ }\bibfield  {title}
  {\enquote {\bibinfo {title} {{Analytical macrospin modeling of the stochastic
  switching time of spin-transfer torque devices}},}\ }\href {\doibase
  10.1109/TED.2014.2372475} {\bibfield  {journal} {\bibinfo  {journal} {IEEE
  Transactions on Electron Devices}\ }\textbf {\bibinfo {volume} {62}},\
  \bibinfo {pages} {164--170} (\bibinfo {year} {2015})}\BibitemShut {NoStop}%
\bibitem [{\citenamefont {Moon}, \citenamefont {Lee},\ and\ \citenamefont
  {You}(2018)}]{Moon2018}%
  \BibitemOpen
  \bibfield  {author} {\bibinfo {author} {\bibfnamefont {J.-H.}\ \bibnamefont
  {Moon}}, \bibinfo {author} {\bibfnamefont {T.~Y.}\ \bibnamefont {Lee}}, \
  and\ \bibinfo {author} {\bibfnamefont {C.-Y.}\ \bibnamefont {You}},\
  }\bibfield  {title} {\enquote {\bibinfo {title} {{Relation between switching
  time distribution and damping constant in magnetic nanostructure}},}\ }\href
  {\doibase 10.1038/s41598-018-31299-4} {\bibfield  {journal} {\bibinfo
  {journal} {Scientific Reports}\ }\textbf {\bibinfo {volume} {8}},\ \bibinfo
  {pages} {13288} (\bibinfo {year} {2018})}\BibitemShut {NoStop}%
\bibitem [{\citenamefont {Siracusano}\ \emph {et~al.}(2018)\citenamefont
  {Siracusano}, \citenamefont {Tomasello}, \citenamefont {D'Aquino},
  \citenamefont {Puliafito}, \citenamefont {Giordano}, \citenamefont
  {Azzerboni}, \citenamefont {Braganca}, \citenamefont {Finocchio},\ and\
  \citenamefont {Carpentieri}}]{Siracusano2018}%
  \BibitemOpen
  \bibfield  {author} {\bibinfo {author} {\bibfnamefont {G.}~\bibnamefont
  {Siracusano}}, \bibinfo {author} {\bibfnamefont {R.}~\bibnamefont
  {Tomasello}}, \bibinfo {author} {\bibfnamefont {M.}~\bibnamefont {D'Aquino}},
  \bibinfo {author} {\bibfnamefont {V.}~\bibnamefont {Puliafito}}, \bibinfo
  {author} {\bibfnamefont {A.}~\bibnamefont {Giordano}}, \bibinfo {author}
  {\bibfnamefont {B.}~\bibnamefont {Azzerboni}}, \bibinfo {author}
  {\bibfnamefont {P.}~\bibnamefont {Braganca}}, \bibinfo {author}
  {\bibfnamefont {G.}~\bibnamefont {Finocchio}}, \ and\ \bibinfo {author}
  {\bibfnamefont {M.}~\bibnamefont {Carpentieri}},\ }\bibfield  {title}
  {\enquote {\bibinfo {title} {{Description of Statistical Switching in
  Perpendicular STT-MRAM Within an Analytical and Numerical Micromagnetic
  Framework}},}\ }\href {\doibase 10.1109/TMAG.2018.2799856} {\bibfield
  {journal} {\bibinfo  {journal} {IEEE Transactions on Magnetics}\ }\textbf
  {\bibinfo {volume} {54}},\ \bibinfo {pages} {1--10} (\bibinfo {year}
  {2018})},\ \Eprint {http://arxiv.org/abs/1702.07739} {arXiv:1702.07739}
  \BibitemShut {NoStop}%
\bibitem [{\citenamefont {D'Aquino}, \citenamefont {Perna},\ and\ \citenamefont
  {Serpico}(2020)}]{DAquino2020}%
  \BibitemOpen
  \bibfield  {author} {\bibinfo {author} {\bibfnamefont {M.}~\bibnamefont
  {D'Aquino}}, \bibinfo {author} {\bibfnamefont {S.}~\bibnamefont {Perna}}, \
  and\ \bibinfo {author} {\bibfnamefont {C.}~\bibnamefont {Serpico}},\
  }\bibfield  {title} {\enquote {\bibinfo {title} {{Micromagnetic study of
  statistical switching in magnetic tunnel junctions stabilized by
  perpendicular shape anisotropy}},}\ }\href {\doibase
  10.1016/j.physb.2019.411744} {\bibfield  {journal} {\bibinfo  {journal}
  {Physica B: Condensed Matter}\ }\textbf {\bibinfo {volume} {577}},\ \bibinfo
  {pages} {411744} (\bibinfo {year} {2020})}\BibitemShut {NoStop}%
\bibitem [{\citenamefont {Shukla}, \citenamefont {Parthasarathy},\ and\
  \citenamefont {Rakheja}(2020)}]{Shukla2020}%
  \BibitemOpen
  \bibfield  {author} {\bibinfo {author} {\bibfnamefont {A.}~\bibnamefont
  {Shukla}}, \bibinfo {author} {\bibfnamefont {A.}~\bibnamefont
  {Parthasarathy}}, \ and\ \bibinfo {author} {\bibfnamefont {S.}~\bibnamefont
  {Rakheja}},\ }\bibfield  {title} {\enquote {\bibinfo {title} {{Switching Time
  of Spin-Torque-Driven Magnetization in Biaxial Ferromagnets}},}\ }\href
  {\doibase 10.1103/physrevapplied.13.054020} {\bibfield  {journal} {\bibinfo
  {journal} {Physical Review Applied}\ }\textbf {\bibinfo {volume} {13}},\
  \bibinfo {pages} {1} (\bibinfo {year} {2020})}\BibitemShut {NoStop}%
\bibitem [{\citenamefont {Mimura}\ \emph {et~al.}(1978)\citenamefont {Mimura},
  \citenamefont {Imamura}, \citenamefont {Kobayashi}, \citenamefont {Okada},\
  and\ \citenamefont {Kushiro}}]{Mimura1978}%
  \BibitemOpen
  \bibfield  {author} {\bibinfo {author} {\bibfnamefont {Y.}~\bibnamefont
  {Mimura}}, \bibinfo {author} {\bibfnamefont {N.}~\bibnamefont {Imamura}},
  \bibinfo {author} {\bibfnamefont {T.}~\bibnamefont {Kobayashi}}, \bibinfo
  {author} {\bibfnamefont {A.}~\bibnamefont {Okada}}, \ and\ \bibinfo {author}
  {\bibfnamefont {Y.}~\bibnamefont {Kushiro}},\ }\bibfield  {title} {\enquote
  {\bibinfo {title} {{Magnetic properties of amorphous alloy films of Fe with
  Gd, Tb, Dy, Ho, or Er}},}\ }\href {\doibase 10.1063/1.325008} {\bibfield
  {journal} {\bibinfo  {journal} {Journal of Applied Physics}\ }\textbf
  {\bibinfo {volume} {49}},\ \bibinfo {pages} {1208--1215} (\bibinfo {year}
  {1978})}\BibitemShut {NoStop}%
\bibitem [{\citenamefont {Krishnaswamy}\ \emph {et~al.}(2022)\citenamefont
  {Krishnaswamy}, \citenamefont {Sala}, \citenamefont {Jacot}, \citenamefont
  {Lambert}, \citenamefont {Schlitz}, \citenamefont {Rossell}, \citenamefont
  {N{\"{o}}el},\ and\ \citenamefont {Gambardella}}]{Krishnaswamy2022}%
  \BibitemOpen
  \bibfield  {author} {\bibinfo {author} {\bibfnamefont {G.~K.}\ \bibnamefont
  {Krishnaswamy}}, \bibinfo {author} {\bibfnamefont {G.}~\bibnamefont {Sala}},
  \bibinfo {author} {\bibfnamefont {B.}~\bibnamefont {Jacot}}, \bibinfo
  {author} {\bibfnamefont {C.-H.}\ \bibnamefont {Lambert}}, \bibinfo {author}
  {\bibfnamefont {R.}~\bibnamefont {Schlitz}}, \bibinfo {author} {\bibfnamefont
  {M.~D.}\ \bibnamefont {Rossell}}, \bibinfo {author} {\bibfnamefont
  {P.}~\bibnamefont {N{\"{o}}el}}, \ and\ \bibinfo {author} {\bibfnamefont
  {P.}~\bibnamefont {Gambardella}},\ }\bibfield  {title} {\enquote {\bibinfo
  {title} {{Time-Dependent Multistate Switching of Topological
  Antiferromagnetic Order in Mn3Sn}},}\ }\href {\doibase
  10.1103/PhysRevApplied.18.024064} {\bibfield  {journal} {\bibinfo  {journal}
  {Physical Review Applied}\ }\textbf {\bibinfo {volume} {18}},\ \bibinfo
  {pages} {024064} (\bibinfo {year} {2022})},\ \Eprint
  {http://arxiv.org/abs/2205.05309} {arXiv:2205.05309} \BibitemShut {NoStop}%
\bibitem [{\citenamefont {Sala}\ and\ \citenamefont
  {Gambardella}(2022)}]{Sala2022b}%
  \BibitemOpen
  \bibfield  {author} {\bibinfo {author} {\bibfnamefont {G.}~\bibnamefont
  {Sala}}\ and\ \bibinfo {author} {\bibfnamefont {P.}~\bibnamefont
  {Gambardella}},\ }\bibfield  {title} {\enquote {\bibinfo {title}
  {{Ferrimagnetic Dynamics Induced by Spin‐Orbit Torques}},}\ }\href
  {\doibase 10.1002/admi.202201622} {\bibfield  {journal} {\bibinfo  {journal}
  {Advanced Materials Interfaces}\ }\textbf {\bibinfo {volume} {2201622}}
  (\bibinfo {year} {2022}),\ 10.1002/admi.202201622}\BibitemShut {NoStop}%
\bibitem [{\citenamefont {Haazen}\ \emph {et~al.}(2013)\citenamefont {Haazen},
  \citenamefont {Mur{\`{e}}}, \citenamefont {Franken}, \citenamefont
  {Lavrijsen}, \citenamefont {Swagten},\ and\ \citenamefont
  {Koopmans}}]{Haazen2013}%
  \BibitemOpen
  \bibfield  {author} {\bibinfo {author} {\bibfnamefont {P.~P.~J.}\
  \bibnamefont {Haazen}}, \bibinfo {author} {\bibfnamefont {E.}~\bibnamefont
  {Mur{\`{e}}}}, \bibinfo {author} {\bibfnamefont {J.~H.}\ \bibnamefont
  {Franken}}, \bibinfo {author} {\bibfnamefont {R.}~\bibnamefont {Lavrijsen}},
  \bibinfo {author} {\bibfnamefont {H.~J.~M.}\ \bibnamefont {Swagten}}, \ and\
  \bibinfo {author} {\bibfnamefont {B.}~\bibnamefont {Koopmans}},\ }\bibfield
  {title} {\enquote {\bibinfo {title} {{Domain wall depinning governed by the
  spin Hall effect}},}\ }\href {\doibase 10.1038/nmat3553} {\bibfield
  {journal} {\bibinfo  {journal} {Nature Materials}\ }\textbf {\bibinfo
  {volume} {12}},\ \bibinfo {pages} {299--303} (\bibinfo {year}
  {2013})}\BibitemShut {NoStop}%
\bibitem [{\citenamefont {Emori}\ \emph {et~al.}(2013)\citenamefont {Emori},
  \citenamefont {Bauer}, \citenamefont {Ahn}, \citenamefont {Martinez},\ and\
  \citenamefont {Beach}}]{Emori2013}%
  \BibitemOpen
  \bibfield  {author} {\bibinfo {author} {\bibfnamefont {S.}~\bibnamefont
  {Emori}}, \bibinfo {author} {\bibfnamefont {U.}~\bibnamefont {Bauer}},
  \bibinfo {author} {\bibfnamefont {S.-M.}\ \bibnamefont {Ahn}}, \bibinfo
  {author} {\bibfnamefont {E.}~\bibnamefont {Martinez}}, \ and\ \bibinfo
  {author} {\bibfnamefont {G.~S.~D.}\ \bibnamefont {Beach}},\ }\bibfield
  {title} {\enquote {\bibinfo {title} {{Current-driven dynamics of chiral
  ferromagnetic domain walls}},}\ }\href {\doibase 10.1038/nmat3675} {\bibfield
   {journal} {\bibinfo  {journal} {Nature Materials}\ }\textbf {\bibinfo
  {volume} {12}},\ \bibinfo {pages} {611--616} (\bibinfo {year} {2013})},\
  \Eprint {http://arxiv.org/abs/1302.2257} {arXiv:1302.2257} \BibitemShut
  {NoStop}%
\bibitem [{\citenamefont {Emori}\ \emph {et~al.}(2014)\citenamefont {Emori},
  \citenamefont {Martinez}, \citenamefont {Lee}, \citenamefont {Lee},
  \citenamefont {Bauer}, \citenamefont {Ahn}, \citenamefont {Agrawal},
  \citenamefont {Bono},\ and\ \citenamefont {Beach}}]{Emori2014}%
  \BibitemOpen
  \bibfield  {author} {\bibinfo {author} {\bibfnamefont {S.}~\bibnamefont
  {Emori}}, \bibinfo {author} {\bibfnamefont {E.}~\bibnamefont {Martinez}},
  \bibinfo {author} {\bibfnamefont {K.-J.}\ \bibnamefont {Lee}}, \bibinfo
  {author} {\bibfnamefont {H.-W.}\ \bibnamefont {Lee}}, \bibinfo {author}
  {\bibfnamefont {U.}~\bibnamefont {Bauer}}, \bibinfo {author} {\bibfnamefont
  {S.-M.}\ \bibnamefont {Ahn}}, \bibinfo {author} {\bibfnamefont
  {P.}~\bibnamefont {Agrawal}}, \bibinfo {author} {\bibfnamefont {D.~C.}\
  \bibnamefont {Bono}}, \ and\ \bibinfo {author} {\bibfnamefont {G.~S.~D.}\
  \bibnamefont {Beach}},\ }\bibfield  {title} {\enquote {\bibinfo {title}
  {{Spin Hall torque magnetometry of Dzyaloshinskii domain walls}},}\ }\href
  {\doibase 10.1103/PhysRevB.90.184427} {\bibfield  {journal} {\bibinfo
  {journal} {Physical Review B}\ }\textbf {\bibinfo {volume} {90}},\ \bibinfo
  {pages} {184427} (\bibinfo {year} {2014})}\BibitemShut {NoStop}%
\bibitem [{\citenamefont {Martinez}\ \emph {et~al.}(2014)\citenamefont
  {Martinez}, \citenamefont {Emori}, \citenamefont {Perez}, \citenamefont
  {Torres},\ and\ \citenamefont {Beach}}]{Martinez2014}%
  \BibitemOpen
  \bibfield  {author} {\bibinfo {author} {\bibfnamefont {E.}~\bibnamefont
  {Martinez}}, \bibinfo {author} {\bibfnamefont {S.}~\bibnamefont {Emori}},
  \bibinfo {author} {\bibfnamefont {N.}~\bibnamefont {Perez}}, \bibinfo
  {author} {\bibfnamefont {L.}~\bibnamefont {Torres}}, \ and\ \bibinfo {author}
  {\bibfnamefont {G.~S.~D.}\ \bibnamefont {Beach}},\ }\bibfield  {title}
  {\enquote {\bibinfo {title} {{Current-driven dynamics of Dzyaloshinskii
  domain walls in the presence of in-plane fields: Full micromagnetic and
  one-dimensional analysis}},}\ }\href {\doibase 10.1063/1.4881778} {\bibfield
  {journal} {\bibinfo  {journal} {Journal of Applied Physics}\ }\textbf
  {\bibinfo {volume} {115}},\ \bibinfo {pages} {213909} (\bibinfo {year}
  {2014})}\BibitemShut {NoStop}%
\bibitem [{\citenamefont {Yang}, \citenamefont {Ryu},\ and\ \citenamefont
  {Parkin}(2015)}]{Yang2015}%
  \BibitemOpen
  \bibfield  {author} {\bibinfo {author} {\bibfnamefont {S.-H.}\ \bibnamefont
  {Yang}}, \bibinfo {author} {\bibfnamefont {K.-S.}\ \bibnamefont {Ryu}}, \
  and\ \bibinfo {author} {\bibfnamefont {S.}~\bibnamefont {Parkin}},\
  }\bibfield  {title} {\enquote {\bibinfo {title} {{Domain-wall velocities of
  up to 750 m s-1 driven by exchange-coupling torque in synthetic
  antiferromagnets}},}\ }\href {\doibase 10.1038/nnano.2014.324} {\bibfield
  {journal} {\bibinfo  {journal} {Nature Nanotechnology}\ }\textbf {\bibinfo
  {volume} {10}},\ \bibinfo {pages} {221--226} (\bibinfo {year}
  {2015})}\BibitemShut {NoStop}%
\bibitem [{\citenamefont {Lee}\ \emph {et~al.}(2014)\citenamefont {Lee},
  \citenamefont {Lee}, \citenamefont {Min},\ and\ \citenamefont
  {Lee}}]{Lee2014}%
  \BibitemOpen
  \bibfield  {author} {\bibinfo {author} {\bibfnamefont {K.-S.}\ \bibnamefont
  {Lee}}, \bibinfo {author} {\bibfnamefont {S.-W.}\ \bibnamefont {Lee}},
  \bibinfo {author} {\bibfnamefont {B.-C.}\ \bibnamefont {Min}}, \ and\
  \bibinfo {author} {\bibfnamefont {K.-J.}\ \bibnamefont {Lee}},\ }\bibfield
  {title} {\enquote {\bibinfo {title} {{Thermally activated switching of
  perpendicular magnet by spin-orbit spin torque}},}\ }\href {\doibase
  10.1063/1.4866186} {\bibfield  {journal} {\bibinfo  {journal} {Applied
  Physics Letters}\ }\textbf {\bibinfo {volume} {104}},\ \bibinfo {pages}
  {072413} (\bibinfo {year} {2014})}\BibitemShut {NoStop}%
\bibitem [{\citenamefont {{Lo Conte}}\ \emph {et~al.}(2014)\citenamefont {{Lo
  Conte}}, \citenamefont {Hrabec}, \citenamefont {Mihai}, \citenamefont
  {Schulz}, \citenamefont {Noh}, \citenamefont {Marrows}, \citenamefont
  {Moore},\ and\ \citenamefont {Kl{\"{a}}ui}}]{LoConte2014}%
  \BibitemOpen
  \bibfield  {author} {\bibinfo {author} {\bibfnamefont {R.}~\bibnamefont {{Lo
  Conte}}}, \bibinfo {author} {\bibfnamefont {A.}~\bibnamefont {Hrabec}},
  \bibinfo {author} {\bibfnamefont {A.~P.}\ \bibnamefont {Mihai}}, \bibinfo
  {author} {\bibfnamefont {T.}~\bibnamefont {Schulz}}, \bibinfo {author}
  {\bibfnamefont {S.~J.}\ \bibnamefont {Noh}}, \bibinfo {author} {\bibfnamefont
  {C.~H.}\ \bibnamefont {Marrows}}, \bibinfo {author} {\bibfnamefont {T.~A.}\
  \bibnamefont {Moore}}, \ and\ \bibinfo {author} {\bibfnamefont
  {M.}~\bibnamefont {Kl{\"{a}}ui}},\ }\bibfield  {title} {\enquote {\bibinfo
  {title} {{Spin-orbit torque-driven magnetization switching and thermal
  effects studied in Ta$\backslash$CoFeB$\backslash$MgO nanowires}},}\ }\href
  {\doibase 10.1063/1.4896225} {\bibfield  {journal} {\bibinfo  {journal}
  {Applied Physics Letters}\ }\textbf {\bibinfo {volume} {105}} (\bibinfo
  {year} {2014}),\ 10.1063/1.4896225},\ \Eprint
  {http://arxiv.org/abs/1405.0452} {arXiv:1405.0452} \BibitemShut {NoStop}%
\bibitem [{\citenamefont {Kim}\ \emph {et~al.}(2017)\citenamefont {Kim},
  \citenamefont {Jang}, \citenamefont {Kim}, \citenamefont {Ishibashi},
  \citenamefont {Taniguchi}, \citenamefont {Moriyama}, \citenamefont {Kim},
  \citenamefont {Lee},\ and\ \citenamefont {Ono}}]{Kim2017d}%
  \BibitemOpen
  \bibfield  {author} {\bibinfo {author} {\bibfnamefont {S.}~\bibnamefont
  {Kim}}, \bibinfo {author} {\bibfnamefont {P.-h.}\ \bibnamefont {Jang}},
  \bibinfo {author} {\bibfnamefont {D.-h.}\ \bibnamefont {Kim}}, \bibinfo
  {author} {\bibfnamefont {M.}~\bibnamefont {Ishibashi}}, \bibinfo {author}
  {\bibfnamefont {T.}~\bibnamefont {Taniguchi}}, \bibinfo {author}
  {\bibfnamefont {T.}~\bibnamefont {Moriyama}}, \bibinfo {author}
  {\bibfnamefont {K.-j.}\ \bibnamefont {Kim}}, \bibinfo {author} {\bibfnamefont
  {K.-j.}\ \bibnamefont {Lee}}, \ and\ \bibinfo {author} {\bibfnamefont
  {T.}~\bibnamefont {Ono}},\ }\bibfield  {title} {\enquote {\bibinfo {title}
  {{Magnetic droplet nucleation with a homochiral N{\'{e}}el domain wall}},}\
  }\href {\doibase 10.1103/PhysRevB.95.220402} {\bibfield  {journal} {\bibinfo
  {journal} {Physical Review B}\ }\textbf {\bibinfo {volume} {95}},\ \bibinfo
  {pages} {220402} (\bibinfo {year} {2017})}\BibitemShut {NoStop}%
\bibitem [{\citenamefont {Brown}(1963)}]{Brown1963}%
  \BibitemOpen
  \bibfield  {author} {\bibinfo {author} {\bibfnamefont {W.~F.}\ \bibnamefont
  {Brown}},\ }\bibfield  {title} {\enquote {\bibinfo {title} {{Thermal
  Fluctuations of a Single‐Domain Particle}},}\ }\href {\doibase
  10.1063/1.1729489} {\bibfield  {journal} {\bibinfo  {journal} {Journal of
  Applied Physics}\ }\textbf {\bibinfo {volume} {34}},\ \bibinfo {pages}
  {1319--1320} (\bibinfo {year} {1963})}\BibitemShut {NoStop}%
\bibitem [{\citenamefont {Garc{\'{i}}a-Palacios}\ and\ \citenamefont
  {L{\'{a}}zaro}(1998)}]{Garcia1998}%
  \BibitemOpen
  \bibfield  {author} {\bibinfo {author} {\bibfnamefont {J.~L.}\ \bibnamefont
  {Garc{\'{i}}a-Palacios}}\ and\ \bibinfo {author} {\bibfnamefont {F.~J.}\
  \bibnamefont {L{\'{a}}zaro}},\ }\bibfield  {title} {\enquote {\bibinfo
  {title} {{Langevin-dynamics study of the dynamical properties of small
  magnetic particles}},}\ }\href {\doibase 10.1103/PhysRevB.58.14937}
  {\bibfield  {journal} {\bibinfo  {journal} {Physical Review B}\ }\textbf
  {\bibinfo {volume} {58}},\ \bibinfo {pages} {14937--14958} (\bibinfo {year}
  {1998})}\BibitemShut {NoStop}%
\bibitem [{\citenamefont {Xian-Ying}\ \emph {et~al.}(2003)\citenamefont
  {Xian-Ying}, \citenamefont {Yue-Pin}, \citenamefont {Zuo-Yi}, \citenamefont
  {De-Fang},\ and\ \citenamefont {Fu-Xi}}]{Xian-Ying2003}%
  \BibitemOpen
  \bibfield  {author} {\bibinfo {author} {\bibfnamefont {W.}~\bibnamefont
  {Xian-Ying}}, \bibinfo {author} {\bibfnamefont {Z.}~\bibnamefont {Yue-Pin}},
  \bibinfo {author} {\bibfnamefont {L.}~\bibnamefont {Zuo-Yi}}, \bibinfo
  {author} {\bibfnamefont {S.}~\bibnamefont {De-Fang}}, \ and\ \bibinfo
  {author} {\bibfnamefont {G.}~\bibnamefont {Fu-Xi}},\ }\bibfield  {title}
  {\enquote {\bibinfo {title} {{Temperature-Induced Magnetization Reorientation
  in GdFeCo/ TbFeCo Exchange-Coupled Double Layer Films}},}\ }\href {\doibase
  10.1088/0256-307x/20/8/352} {\bibfield  {journal} {\bibinfo  {journal}
  {Chinese Physics Letters}\ }\textbf {\bibinfo {volume} {20}},\ \bibinfo
  {pages} {1359--1361} (\bibinfo {year} {2003})}\BibitemShut {NoStop}%
\bibitem [{\citenamefont {Raasch}\ \emph {et~al.}(1994)\citenamefont {Raasch},
  \citenamefont {Reck}, \citenamefont {Mathieu},\ and\ \citenamefont
  {Hillebrands}}]{Raasch1994}%
  \BibitemOpen
  \bibfield  {author} {\bibinfo {author} {\bibfnamefont {D.}~\bibnamefont
  {Raasch}}, \bibinfo {author} {\bibfnamefont {J.}~\bibnamefont {Reck}},
  \bibinfo {author} {\bibfnamefont {C.}~\bibnamefont {Mathieu}}, \ and\
  \bibinfo {author} {\bibfnamefont {B.}~\bibnamefont {Hillebrands}},\
  }\bibfield  {title} {\enquote {\bibinfo {title} {{Exchange stiffness constant
  and wall energy density of amorphous GdTb-FeCo thin films}},}\ }\href
  {\doibase 10.1063/1.357837} {\bibfield  {journal} {\bibinfo  {journal}
  {Journal of Applied Physics}\ }\textbf {\bibinfo {volume} {76}},\ \bibinfo
  {pages} {1145--1149} (\bibinfo {year} {1994})}\BibitemShut {NoStop}%
\bibitem [{\citenamefont {Radu}\ and\ \citenamefont
  {S{\'{a}}nchez-Barriga}(2018)}]{Radu2018}%
  \BibitemOpen
  \bibfield  {author} {\bibinfo {author} {\bibfnamefont {F.}~\bibnamefont
  {Radu}}\ and\ \bibinfo {author} {\bibfnamefont {J.}~\bibnamefont
  {S{\'{a}}nchez-Barriga}},\ }\bibfield  {title} {\enquote {\bibinfo {title}
  {{Ferrimagnetic Heterostructures for Applications in Magnetic Recording}},}\
  }in\ \href {\doibase 10.1016/B978-0-12-813594-5.00009-6} {\emph {\bibinfo
  {booktitle} {Novel Magnetic Nanostructures}}}\ (\bibinfo  {publisher}
  {Elsevier},\ \bibinfo {year} {2018})\ pp.\ \bibinfo {pages}
  {267--331}\BibitemShut {NoStop}%
\bibitem [{\citenamefont {Finley}\ and\ \citenamefont
  {Liu}(2016)}]{Finley2016}%
  \BibitemOpen
  \bibfield  {author} {\bibinfo {author} {\bibfnamefont {J.}~\bibnamefont
  {Finley}}\ and\ \bibinfo {author} {\bibfnamefont {L.}~\bibnamefont {Liu}},\
  }\bibfield  {title} {\enquote {\bibinfo {title} {{Spin-Orbit-Torque
  Efficiency in Compensated Ferrimagnetic Cobalt-Terbium Alloys}},}\ }\href
  {\doibase 10.1103/PhysRevApplied.6.054001} {\bibfield  {journal} {\bibinfo
  {journal} {Physical Review Applied}\ }\textbf {\bibinfo {volume} {6}},\
  \bibinfo {pages} {054001} (\bibinfo {year} {2016})}\BibitemShut {NoStop}%
\bibitem [{\citenamefont {Shahbazi}\ \emph {et~al.}(2019)\citenamefont
  {Shahbazi}, \citenamefont {Kim}, \citenamefont {Nembach}, \citenamefont
  {Shaw}, \citenamefont {Bischof}, \citenamefont {Rossell}, \citenamefont
  {Jeudy}, \citenamefont {Moore},\ and\ \citenamefont
  {Marrows}}]{Shahbazi2019}%
  \BibitemOpen
  \bibfield  {author} {\bibinfo {author} {\bibfnamefont {K.}~\bibnamefont
  {Shahbazi}}, \bibinfo {author} {\bibfnamefont {J.-V.}\ \bibnamefont {Kim}},
  \bibinfo {author} {\bibfnamefont {H.~T.}\ \bibnamefont {Nembach}}, \bibinfo
  {author} {\bibfnamefont {J.~M.}\ \bibnamefont {Shaw}}, \bibinfo {author}
  {\bibfnamefont {A.}~\bibnamefont {Bischof}}, \bibinfo {author} {\bibfnamefont
  {M.~D.}\ \bibnamefont {Rossell}}, \bibinfo {author} {\bibfnamefont
  {V.}~\bibnamefont {Jeudy}}, \bibinfo {author} {\bibfnamefont {T.~A.}\
  \bibnamefont {Moore}}, \ and\ \bibinfo {author} {\bibfnamefont {C.~H.}\
  \bibnamefont {Marrows}},\ }\bibfield  {title} {\enquote {\bibinfo {title}
  {{Domain-wall motion and interfacial Dzyaloshinskii-Moriya interactions
  Pt/Co/Ir(t)/Ta multilayers}},}\ }\href {\doibase 10.1103/PhysRevB.99.094409}
  {\bibfield  {journal} {\bibinfo  {journal} {Physical Review B}\ }\textbf
  {\bibinfo {volume} {99}},\ \bibinfo {pages} {094409} (\bibinfo {year}
  {2019})}\BibitemShut {NoStop}%
\bibitem [{\citenamefont {Kim}\ \emph {et~al.}(2019)\citenamefont {Kim},
  \citenamefont {Okuno}, \citenamefont {Kim}, \citenamefont {Oh}, \citenamefont
  {Nishimura}, \citenamefont {Hirata}, \citenamefont {Futakawa}, \citenamefont
  {Yoshikawa}, \citenamefont {Tsukamoto}, \citenamefont {Tserkovnyak},
  \citenamefont {Shiota}, \citenamefont {Moriyama}, \citenamefont {Kim},
  \citenamefont {Lee},\ and\ \citenamefont {Ono}}]{Kim2019}%
  \BibitemOpen
  \bibfield  {author} {\bibinfo {author} {\bibfnamefont {D.-H.}\ \bibnamefont
  {Kim}}, \bibinfo {author} {\bibfnamefont {T.}~\bibnamefont {Okuno}}, \bibinfo
  {author} {\bibfnamefont {S.~K.}\ \bibnamefont {Kim}}, \bibinfo {author}
  {\bibfnamefont {S.-H.}\ \bibnamefont {Oh}}, \bibinfo {author} {\bibfnamefont
  {T.}~\bibnamefont {Nishimura}}, \bibinfo {author} {\bibfnamefont
  {Y.}~\bibnamefont {Hirata}}, \bibinfo {author} {\bibfnamefont
  {Y.}~\bibnamefont {Futakawa}}, \bibinfo {author} {\bibfnamefont
  {H.}~\bibnamefont {Yoshikawa}}, \bibinfo {author} {\bibfnamefont
  {A.}~\bibnamefont {Tsukamoto}}, \bibinfo {author} {\bibfnamefont
  {Y.}~\bibnamefont {Tserkovnyak}}, \bibinfo {author} {\bibfnamefont
  {Y.}~\bibnamefont {Shiota}}, \bibinfo {author} {\bibfnamefont
  {T.}~\bibnamefont {Moriyama}}, \bibinfo {author} {\bibfnamefont {K.-J.}\
  \bibnamefont {Kim}}, \bibinfo {author} {\bibfnamefont {K.-J.}\ \bibnamefont
  {Lee}}, \ and\ \bibinfo {author} {\bibfnamefont {T.}~\bibnamefont {Ono}},\
  }\bibfield  {title} {\enquote {\bibinfo {title} {{Low Magnetic Damping of
  Ferrimagnetic GdFeCo Alloys}},}\ }\href {\doibase
  10.1103/PhysRevLett.122.127203} {\bibfield  {journal} {\bibinfo  {journal}
  {Physical Review Letters}\ }\textbf {\bibinfo {volume} {122}},\ \bibinfo
  {pages} {127203} (\bibinfo {year} {2019})}\BibitemShut {NoStop}%
\bibitem [{\citenamefont {Mougin}\ \emph {et~al.}(2007)\citenamefont {Mougin},
  \citenamefont {Cormier}, \citenamefont {Adam}, \citenamefont {Metaxas},\ and\
  \citenamefont {Ferr{\'{e}}}}]{Mougin2007}%
  \BibitemOpen
  \bibfield  {author} {\bibinfo {author} {\bibfnamefont {A.}~\bibnamefont
  {Mougin}}, \bibinfo {author} {\bibfnamefont {M.}~\bibnamefont {Cormier}},
  \bibinfo {author} {\bibfnamefont {J.~P.}\ \bibnamefont {Adam}}, \bibinfo
  {author} {\bibfnamefont {P.~J.}\ \bibnamefont {Metaxas}}, \ and\ \bibinfo
  {author} {\bibfnamefont {J.}~\bibnamefont {Ferr{\'{e}}}},\ }\bibfield
  {title} {\enquote {\bibinfo {title} {{Domain wall mobility, stability and
  Walker breakdown in magnetic nanowires}},}\ }\href {\doibase
  10.1209/0295-5075/78/57007} {\bibfield  {journal} {\bibinfo  {journal}
  {Europhysics Letters (EPL)}\ }\textbf {\bibinfo {volume} {78}},\ \bibinfo
  {pages} {57007} (\bibinfo {year} {2007})},\ \Eprint
  {http://arxiv.org/abs/0702492} {arXiv:0702492 [cond-mat]} \BibitemShut
  {NoStop}%
\bibitem [{\citenamefont {Wernsdorfer}\ \emph {et~al.}(1997)\citenamefont
  {Wernsdorfer}, \citenamefont {Hasselbach}, \citenamefont {Benoit},
  \citenamefont {Barbara}, \citenamefont {Doudin}, \citenamefont {Meier},
  \citenamefont {Ansermet},\ and\ \citenamefont {Mailly}}]{Wernsdorfer1997}%
  \BibitemOpen
  \bibfield  {author} {\bibinfo {author} {\bibfnamefont {W.}~\bibnamefont
  {Wernsdorfer}}, \bibinfo {author} {\bibfnamefont {K.}~\bibnamefont
  {Hasselbach}}, \bibinfo {author} {\bibfnamefont {A.}~\bibnamefont {Benoit}},
  \bibinfo {author} {\bibfnamefont {B.}~\bibnamefont {Barbara}}, \bibinfo
  {author} {\bibfnamefont {B.}~\bibnamefont {Doudin}}, \bibinfo {author}
  {\bibfnamefont {J.}~\bibnamefont {Meier}}, \bibinfo {author} {\bibfnamefont
  {J.-P.}\ \bibnamefont {Ansermet}}, \ and\ \bibinfo {author} {\bibfnamefont
  {D.}~\bibnamefont {Mailly}},\ }\bibfield  {title} {\enquote {\bibinfo {title}
  {{Measurements of magnetization switching in individual nickel nanowires}},}\
  }\href {\doibase 10.1103/PhysRevB.55.11552} {\bibfield  {journal} {\bibinfo
  {journal} {Physical Review B}\ }\textbf {\bibinfo {volume} {55}},\ \bibinfo
  {pages} {11552--11559} (\bibinfo {year} {1997})}\BibitemShut {NoStop}%
\bibitem [{\citenamefont {Wuth}, \citenamefont {Lendecke},\ and\ \citenamefont
  {Meier}(2012)}]{Wuth2012}%
  \BibitemOpen
  \bibfield  {author} {\bibinfo {author} {\bibfnamefont {C.}~\bibnamefont
  {Wuth}}, \bibinfo {author} {\bibfnamefont {P.}~\bibnamefont {Lendecke}}, \
  and\ \bibinfo {author} {\bibfnamefont {G.}~\bibnamefont {Meier}},\ }\bibfield
   {title} {\enquote {\bibinfo {title} {{Temperature-dependent dynamics of
  stochastic domain-wall depinning in nanowires}},}\ }\href {\doibase
  10.1088/0953-8984/24/2/024207} {\bibfield  {journal} {\bibinfo  {journal}
  {Journal of Physics: Condensed Matter}\ }\textbf {\bibinfo {volume} {24}},\
  \bibinfo {pages} {024207} (\bibinfo {year} {2012})}\BibitemShut {NoStop}%
\bibitem [{\citenamefont {Mart{\'{i}}nez}, \citenamefont {Raposo},\ and\
  \citenamefont {Alejos}(2019)}]{Martinez2019}%
  \BibitemOpen
  \bibfield  {author} {\bibinfo {author} {\bibfnamefont {E.}~\bibnamefont
  {Mart{\'{i}}nez}}, \bibinfo {author} {\bibfnamefont {V.}~\bibnamefont
  {Raposo}}, \ and\ \bibinfo {author} {\bibfnamefont {{\'{O}}.}~\bibnamefont
  {Alejos}},\ }\bibfield  {title} {\enquote {\bibinfo {title} {{Current-driven
  domain wall dynamics in ferrimagnets: Micromagnetic approach and collective
  coordinates model}},}\ }\href {\doibase 10.1016/j.jmmm.2019.165545}
  {\bibfield  {journal} {\bibinfo  {journal} {Journal of Magnetism and Magnetic
  Materials}\ }\textbf {\bibinfo {volume} {491}} (\bibinfo {year} {2019}),\
  10.1016/j.jmmm.2019.165545},\ \Eprint
  {http://arxiv.org/abs/arXiv:1907.06431v1} {arXiv:arXiv:1907.06431v1}
  \BibitemShut {NoStop}%
\bibitem [{\citenamefont {Moreno}\ \emph {et~al.}(2016)\citenamefont {Moreno},
  \citenamefont {Evans}, \citenamefont {Khmelevskyi}, \citenamefont
  {Mu{\~{n}}oz}, \citenamefont {Chantrell},\ and\ \citenamefont
  {Chubykalo-Fesenko}}]{Moreno2016}%
  \BibitemOpen
  \bibfield  {author} {\bibinfo {author} {\bibfnamefont {R.}~\bibnamefont
  {Moreno}}, \bibinfo {author} {\bibfnamefont {R.~F.~L.}\ \bibnamefont
  {Evans}}, \bibinfo {author} {\bibfnamefont {S.}~\bibnamefont {Khmelevskyi}},
  \bibinfo {author} {\bibfnamefont {M.~C.}\ \bibnamefont {Mu{\~{n}}oz}},
  \bibinfo {author} {\bibfnamefont {R.~W.}\ \bibnamefont {Chantrell}}, \ and\
  \bibinfo {author} {\bibfnamefont {O.}~\bibnamefont {Chubykalo-Fesenko}},\
  }\bibfield  {title} {\enquote {\bibinfo {title} {{Temperature-dependent
  exchange stiffness and domain wall width in Co}},}\ }\href {\doibase
  10.1103/PhysRevB.94.104433} {\bibfield  {journal} {\bibinfo  {journal}
  {Physical Review B}\ }\textbf {\bibinfo {volume} {94}},\ \bibinfo {pages}
  {104433} (\bibinfo {year} {2016})}\BibitemShut {NoStop}%
\bibitem [{\citenamefont {Hirata}\ \emph {et~al.}(2018)\citenamefont {Hirata},
  \citenamefont {Kim}, \citenamefont {Okuno}, \citenamefont {Nishimura},
  \citenamefont {Kim}, \citenamefont {Futakawa}, \citenamefont {Yoshikawa},
  \citenamefont {Tsukamoto}, \citenamefont {Kim}, \citenamefont {Choe},\ and\
  \citenamefont {Ono}}]{Hirata2018}%
  \BibitemOpen
  \bibfield  {author} {\bibinfo {author} {\bibfnamefont {Y.}~\bibnamefont
  {Hirata}}, \bibinfo {author} {\bibfnamefont {D.-H.}\ \bibnamefont {Kim}},
  \bibinfo {author} {\bibfnamefont {T.}~\bibnamefont {Okuno}}, \bibinfo
  {author} {\bibfnamefont {T.}~\bibnamefont {Nishimura}}, \bibinfo {author}
  {\bibfnamefont {D.-Y.}\ \bibnamefont {Kim}}, \bibinfo {author} {\bibfnamefont
  {Y.}~\bibnamefont {Futakawa}}, \bibinfo {author} {\bibfnamefont
  {H.}~\bibnamefont {Yoshikawa}}, \bibinfo {author} {\bibfnamefont
  {A.}~\bibnamefont {Tsukamoto}}, \bibinfo {author} {\bibfnamefont {K.-j.}\
  \bibnamefont {Kim}}, \bibinfo {author} {\bibfnamefont {S.-b.}\ \bibnamefont
  {Choe}}, \ and\ \bibinfo {author} {\bibfnamefont {T.}~\bibnamefont {Ono}},\
  }\bibfield  {title} {\enquote {\bibinfo {title} {{Correlation between
  compensation temperatures of magnetization and angular momentum in GdFeCo
  ferrimagnets}},}\ }\href {\doibase 10.1103/PhysRevB.97.220403} {\bibfield
  {journal} {\bibinfo  {journal} {Physical Review B}\ }\textbf {\bibinfo
  {volume} {97}},\ \bibinfo {pages} {220403} (\bibinfo {year}
  {2018})}\BibitemShut {NoStop}%
\bibitem [{\citenamefont {Martinez}\ \emph {et~al.}(2007)\citenamefont
  {Martinez}, \citenamefont {Lopez-Diaz}, \citenamefont {Alejos}, \citenamefont
  {Torres},\ and\ \citenamefont {Tristan}}]{Martinez2007}%
  \BibitemOpen
  \bibfield  {author} {\bibinfo {author} {\bibfnamefont {E.}~\bibnamefont
  {Martinez}}, \bibinfo {author} {\bibfnamefont {L.}~\bibnamefont
  {Lopez-Diaz}}, \bibinfo {author} {\bibfnamefont {O.}~\bibnamefont {Alejos}},
  \bibinfo {author} {\bibfnamefont {L.}~\bibnamefont {Torres}}, \ and\ \bibinfo
  {author} {\bibfnamefont {C.}~\bibnamefont {Tristan}},\ }\bibfield  {title}
  {\enquote {\bibinfo {title} {{Thermal Effects on Domain Wall Depinning from a
  Single Notch}},}\ }\href {\doibase 10.1103/PhysRevLett.98.267202} {\bibfield
  {journal} {\bibinfo  {journal} {Physical Review Letters}\ }\textbf {\bibinfo
  {volume} {98}},\ \bibinfo {pages} {267202} (\bibinfo {year}
  {2007})}\BibitemShut {NoStop}%
\bibitem [{\citenamefont {Burrowes}\ \emph {et~al.}(2010)\citenamefont
  {Burrowes}, \citenamefont {Mihai}, \citenamefont {Ravelosona}, \citenamefont
  {Kim}, \citenamefont {Chappert}, \citenamefont {Vila}, \citenamefont {Marty},
  \citenamefont {Samson}, \citenamefont {Garcia-Sanchez}, \citenamefont
  {Buda-Prejbeanu}, \citenamefont {Tudosa}, \citenamefont {Fullerton},\ and\
  \citenamefont {Attan{\'{e}}}}]{Burrowes2010}%
  \BibitemOpen
  \bibfield  {author} {\bibinfo {author} {\bibfnamefont {C.}~\bibnamefont
  {Burrowes}}, \bibinfo {author} {\bibfnamefont {A.~P.}\ \bibnamefont {Mihai}},
  \bibinfo {author} {\bibfnamefont {D.}~\bibnamefont {Ravelosona}}, \bibinfo
  {author} {\bibfnamefont {J.-V.}\ \bibnamefont {Kim}}, \bibinfo {author}
  {\bibfnamefont {C.}~\bibnamefont {Chappert}}, \bibinfo {author}
  {\bibfnamefont {L.}~\bibnamefont {Vila}}, \bibinfo {author} {\bibfnamefont
  {A.}~\bibnamefont {Marty}}, \bibinfo {author} {\bibfnamefont
  {Y.}~\bibnamefont {Samson}}, \bibinfo {author} {\bibfnamefont
  {F.}~\bibnamefont {Garcia-Sanchez}}, \bibinfo {author} {\bibfnamefont
  {L.~D.}\ \bibnamefont {Buda-Prejbeanu}}, \bibinfo {author} {\bibfnamefont
  {I.}~\bibnamefont {Tudosa}}, \bibinfo {author} {\bibfnamefont {E.~E.}\
  \bibnamefont {Fullerton}}, \ and\ \bibinfo {author} {\bibfnamefont {J.-P.}\
  \bibnamefont {Attan{\'{e}}}},\ }\bibfield  {title} {\enquote {\bibinfo
  {title} {{Non-adiabatic spin-torques in narrow magnetic domain walls}},}\
  }\href {\doibase 10.1038/nphys1436} {\bibfield  {journal} {\bibinfo
  {journal} {Nature Physics}\ }\textbf {\bibinfo {volume} {6}},\ \bibinfo
  {pages} {17--21} (\bibinfo {year} {2010})}\BibitemShut {NoStop}%
\bibitem [{\citenamefont {Taniguchi}\ \emph {et~al.}(2014)\citenamefont
  {Taniguchi}, \citenamefont {Kim}, \citenamefont {Yoshimura}, \citenamefont
  {Moriyama}, \citenamefont {Tanigawa}, \citenamefont {Suzuki}, \citenamefont
  {Kariyada},\ and\ \citenamefont {Ono}}]{Taniguchi2014}%
  \BibitemOpen
  \bibfield  {author} {\bibinfo {author} {\bibfnamefont {T.}~\bibnamefont
  {Taniguchi}}, \bibinfo {author} {\bibfnamefont {K.-J.}\ \bibnamefont {Kim}},
  \bibinfo {author} {\bibfnamefont {Y.}~\bibnamefont {Yoshimura}}, \bibinfo
  {author} {\bibfnamefont {T.}~\bibnamefont {Moriyama}}, \bibinfo {author}
  {\bibfnamefont {H.}~\bibnamefont {Tanigawa}}, \bibinfo {author}
  {\bibfnamefont {T.}~\bibnamefont {Suzuki}}, \bibinfo {author} {\bibfnamefont
  {E.}~\bibnamefont {Kariyada}}, \ and\ \bibinfo {author} {\bibfnamefont
  {T.}~\bibnamefont {Ono}},\ }\bibfield  {title} {\enquote {\bibinfo {title}
  {{Different stochastic behaviors for magnetic field and current in domain
  wall creep motion}},}\ }\href {\doibase 10.7567/APEX.7.053005} {\bibfield
  {journal} {\bibinfo  {journal} {Applied Physics Express}\ }\textbf {\bibinfo
  {volume} {7}},\ \bibinfo {pages} {053005} (\bibinfo {year}
  {2014})}\BibitemShut {NoStop}%
\bibitem [{\citenamefont {Im}\ \emph {et~al.}(2009)\citenamefont {Im},
  \citenamefont {Bocklage}, \citenamefont {Fischer},\ and\ \citenamefont
  {Meier}}]{Im2009}%
  \BibitemOpen
  \bibfield  {author} {\bibinfo {author} {\bibfnamefont {M.-Y.}\ \bibnamefont
  {Im}}, \bibinfo {author} {\bibfnamefont {L.}~\bibnamefont {Bocklage}},
  \bibinfo {author} {\bibfnamefont {P.}~\bibnamefont {Fischer}}, \ and\
  \bibinfo {author} {\bibfnamefont {G.}~\bibnamefont {Meier}},\ }\bibfield
  {title} {\enquote {\bibinfo {title} {{Direct Observation of Stochastic
  Domain-Wall Depinning in Magnetic Nanowires}},}\ }\href {\doibase
  10.1103/PhysRevLett.102.147204} {\bibfield  {journal} {\bibinfo  {journal}
  {Physical Review Letters}\ }\textbf {\bibinfo {volume} {102}},\ \bibinfo
  {pages} {147204} (\bibinfo {year} {2009})}\BibitemShut {NoStop}%
\bibitem [{\citenamefont {Metaxas}\ \emph {et~al.}(2013)\citenamefont
  {Metaxas}, \citenamefont {Sampaio}, \citenamefont {Chanthbouala},
  \citenamefont {Matsumoto}, \citenamefont {Anane}, \citenamefont {Fert},
  \citenamefont {Zvezdin}, \citenamefont {Yakushiji}, \citenamefont {Kubota},
  \citenamefont {Fukushima}, \citenamefont {Yuasa}, \citenamefont {Nishimura},
  \citenamefont {Nagamine}, \citenamefont {Maehara}, \citenamefont {Tsunekawa},
  \citenamefont {Cros},\ and\ \citenamefont {Grollier}}]{Metaxas2013}%
  \BibitemOpen
  \bibfield  {author} {\bibinfo {author} {\bibfnamefont {P.~J.}\ \bibnamefont
  {Metaxas}}, \bibinfo {author} {\bibfnamefont {J.}~\bibnamefont {Sampaio}},
  \bibinfo {author} {\bibfnamefont {A.}~\bibnamefont {Chanthbouala}}, \bibinfo
  {author} {\bibfnamefont {R.}~\bibnamefont {Matsumoto}}, \bibinfo {author}
  {\bibfnamefont {A.}~\bibnamefont {Anane}}, \bibinfo {author} {\bibfnamefont
  {A.}~\bibnamefont {Fert}}, \bibinfo {author} {\bibfnamefont {K.~A.}\
  \bibnamefont {Zvezdin}}, \bibinfo {author} {\bibfnamefont {K.}~\bibnamefont
  {Yakushiji}}, \bibinfo {author} {\bibfnamefont {H.}~\bibnamefont {Kubota}},
  \bibinfo {author} {\bibfnamefont {A.}~\bibnamefont {Fukushima}}, \bibinfo
  {author} {\bibfnamefont {S.}~\bibnamefont {Yuasa}}, \bibinfo {author}
  {\bibfnamefont {K.}~\bibnamefont {Nishimura}}, \bibinfo {author}
  {\bibfnamefont {Y.}~\bibnamefont {Nagamine}}, \bibinfo {author}
  {\bibfnamefont {H.}~\bibnamefont {Maehara}}, \bibinfo {author} {\bibfnamefont
  {K.}~\bibnamefont {Tsunekawa}}, \bibinfo {author} {\bibfnamefont
  {V.}~\bibnamefont {Cros}}, \ and\ \bibinfo {author} {\bibfnamefont
  {J.}~\bibnamefont {Grollier}},\ }\bibfield  {title} {\enquote {\bibinfo
  {title} {{High domain wall velocities via spin transfer torque using vertical
  current injection}},}\ }\href {\doibase 10.1038/srep01829} {\bibfield
  {journal} {\bibinfo  {journal} {Scientific Reports}\ }\textbf {\bibinfo
  {volume} {3}},\ \bibinfo {pages} {1829} (\bibinfo {year} {2013})}\BibitemShut
  {NoStop}%
\bibitem [{\citenamefont {Attan{\'{e}}}\ \emph {et~al.}(2006)\citenamefont
  {Attan{\'{e}}}, \citenamefont {Ravelosona}, \citenamefont {Marty},
  \citenamefont {Samson},\ and\ \citenamefont {Chappert}}]{Attane2006}%
  \BibitemOpen
  \bibfield  {author} {\bibinfo {author} {\bibfnamefont {J.~P.}\ \bibnamefont
  {Attan{\'{e}}}}, \bibinfo {author} {\bibfnamefont {D.}~\bibnamefont
  {Ravelosona}}, \bibinfo {author} {\bibfnamefont {A.}~\bibnamefont {Marty}},
  \bibinfo {author} {\bibfnamefont {Y.}~\bibnamefont {Samson}}, \ and\ \bibinfo
  {author} {\bibfnamefont {C.}~\bibnamefont {Chappert}},\ }\bibfield  {title}
  {\enquote {\bibinfo {title} {{Thermally Activated Depinning of a Narrow
  Domain Wall from a Single Defect}},}\ }\href {\doibase
  10.1103/PhysRevLett.96.147204} {\bibfield  {journal} {\bibinfo  {journal}
  {Physical Review Letters}\ }\textbf {\bibinfo {volume} {96}},\ \bibinfo
  {pages} {147204} (\bibinfo {year} {2006})}\BibitemShut {NoStop}%
\bibitem [{\citenamefont {Gupta}(1960)}]{Gupta1960}%
  \BibitemOpen
  \bibfield  {author} {\bibinfo {author} {\bibfnamefont {S.~S.}\ \bibnamefont
  {Gupta}},\ }\bibfield  {title} {\enquote {\bibinfo {title} {{Order Statistics
  from the Gamma Distribution}},}\ }\href {\doibase
  10.1080/00401706.1960.10489897} {\bibfield  {journal} {\bibinfo  {journal}
  {Technometrics}\ }\textbf {\bibinfo {volume} {2}},\ \bibinfo {pages}
  {243--262} (\bibinfo {year} {1960})}\BibitemShut {NoStop}%
\bibitem [{\citenamefont {McGill}\ and\ \citenamefont
  {Gibbon}(1965)}]{mcGill1965}%
  \BibitemOpen
  \bibfield  {author} {\bibinfo {author} {\bibfnamefont {W.~J.}\ \bibnamefont
  {McGill}}\ and\ \bibinfo {author} {\bibfnamefont {J.}~\bibnamefont
  {Gibbon}},\ }\bibfield  {title} {\enquote {\bibinfo {title} {{The
  general-gamma distribution and reaction times}},}\ }\href {\doibase
  10.1016/0022-2496(65)90014-3} {\bibfield  {journal} {\bibinfo  {journal}
  {Journal of Mathematical Psychology}\ }\textbf {\bibinfo {volume} {2}},\
  \bibinfo {pages} {1--18} (\bibinfo {year} {1965})}\BibitemShut {NoStop}%
\bibitem [{\citenamefont {Applebaum}(2004)}]{Applebaum2004}%
  \BibitemOpen
  \bibfield  {author} {\bibinfo {author} {\bibfnamefont {D.}~\bibnamefont
  {Applebaum}},\ }\bibfield  {title} {\enquote {\bibinfo {title} {{L{\'{e}}vy
  processes - from probability to finance and quantum groups}},}\ }\href@noop
  {} {\bibfield  {journal} {\bibinfo  {journal} {Notices of the AMS}\ }\textbf
  {\bibinfo {volume} {51}},\ \bibinfo {pages} {1136--1347} (\bibinfo {year}
  {2004})}\BibitemShut {NoStop}%
\end{thebibliography}%

\end{document}